\documentclass[12pt]{article}
\usepackage{graphicx,amsmath,setspace,amssymb}
\usepackage{units}
\usepackage{color}
\usepackage{cite}
\usepackage{hyperref}
\usepackage{multirow}
\usepackage{pbox}
\usepackage{array}
\usepackage{float}
\usepackage{lineno}
\usepackage{hyperref}
\usepackage[caption=true]{subfig}
\usepackage[font=small,labelfont=bf]{caption}

\newlength{\dinwidth}
\newlength{\dinmargin}
\def\lapproxeq{\lower .7ex\hbox{$\;\stackrel{\textstyle                                                    
<}{\sim}\;$}}                                                    
\def\gapproxeq{\lower .7ex\hbox{$\;\stackrel{\textstyle                                                    
>}{\sim}\;$}}                                                    
\def\be{\begin{equation}}                                                    
\def\ee{\end{equation}}                                                    
\def\bea{\begin{eqnarray}}                      
\def\eea{\end{eqnarray}}

\def\GeV{\rm GeV}

\def\sh{\hat s}
\def\sh2{{\hat s}^2}

\begin{document}



\begin{center}

{\Large \bf QED Parton Distribution Functions in the MSHT20 Fit}

\vspace*{1cm}
T. Cridge$^a$, L. A. Harland-Lang$^{b}$, A. D. Martin$^c$, 
and R.S. Thorne$^a$\\                                               
\vspace*{0.5cm}                                                    

$^a$ Department of Physics and Astronomy, University College London, London, WC1E 6BT, UK \\           
$^b$ Rudolf Peierls Centre, Beecroft Building, Parks Road, Oxford, OX1 3PU   \\  
   
$^c$ Institute for Particle Physics Phenomenology, Durham University, Durham, DH1 3LE, UK                   \\

\begin{abstract} 
\noindent We present the MSHT20qed set of parton distribution functions (PDFs). These are obtained from the MSHT20 global analysis via a refit including QED corrections to the DGLAP evolution at ${\cal O}(\alpha),{\cal O}(\alpha\alpha_S)$ and ${\cal O}(\alpha^2)$, and containing the photon PDF of the proton. As in the previous MMHT15qed study we use an input distribution for the photon that is derived from the LUXqed formulation, and find  good consistency for the photon PDF with that of MMHT15qed, as well as with other recent sets. We also present a set of QED corrected neutron PDFs and accompanying photon distribution, and  provide the photon PDF of the nucleons separated into elastic and inelastic contributions. We assess the general expectations for the impact of photon--initiated (PI) corrections to processes entering PDF fits, and review the effect of QED corrections on the other partons and on the fit quality, where electroweak corrections (including PI production) are appropriately added to the cross sections wherever possible. We explore the phenomenological implications of this set by comparing to a variety of benchmark cross sections, finding small but significant corrections due to the inclusion of QED effects in the PDFs. 
\end{abstract}                                                        
\vspace*{0.5cm}

\end{center}

\begin{spacing}{0.8}
\clearpage

\clearpage
\end{spacing}

\section{Introduction  \label{sec:1}} 

The level of  precision aimed for at the Large Hadron Collider (LHC) is now reaching the point where the inclusion of electroweak (EW) corrections in theoretical predictions is often becoming mandatory.  This requires that  EW corrections are applied not only to the partonic cross sections but also to the corresponding 
Parton Distribution Functions (PDFs), with QED corrections forming a part of this. These can be included by supplementing the 
DGLAP~\cite{DGLAP1,DGLAP2,DGLAP3} evolution of the PDFs to include QED parton splittings.
This automatically results in the photon becoming a constituent parton of the proton, leading to photon--initiated (PI) 
sub-processes entering as corrections to the leading QCD cross section for processes such as Drell--Yan \cite{ATLASHMDY8}, 
EW boson--boson scattering \cite{W_W} and Higgs production with an associated EW boson \cite{Higgs+W}. 
Additionally, semi-exclusive \cite{Harland-Lang:2016apc,Harland-Lang:2020veo} and exclusive PI production of states with EW couplings has significant potential as a probe of SM and BSM physics.

The inclusion of QED corrections in DGLAP, and the corresponding photon PDF, has a long history. 
MRST provided the first publicly available QED set\cite{mrstqed}, using DGLAP splitting kernels at $\mathcal{O}(\alpha)$ 
in QED and modelling the input photon as arising radiatively from the quarks below input. Subsequent sets  
either used similar phenomenological models\cite{cteq}, or constrained the photon by utilising the rather limited 
sensitivity of  DIS and Drell-Yan data via PI  final states \cite{nnpdf1,xfitter}. Both approaches lead to photon PDF uncertainties of 
at least $10\%$ and often more.  Moreover, the distinction between the elastic and inelastic photon emission was rarely considered. 
In~\cite{martin_ryskin,Harland-Lang:2016apc,lucian} it was shown how a more accurate determination of the photon distribution at input could be found by using the experimentally 
well determined elastic form factors of the proton. More generally, as discussed long ago in e.g.~\cite{EPA}, the contributions from elastic and inelastic emission to the photon PDF are directly related to the corresponding  structure functions, $F_{1,2}^{ el}$, $F_{1,2}^{inel}$; this idea has been revisited over the years in~\cite{Anlauf:1991wr,Blumlein:1993ef,Mukherjee:2003yh,Luszczak:2015aoa}. However, it has recently been placed within a rigorous and precise theoretical 
framework by the LUXqed group~\cite{LUXQED1,LUXQED2}, who consequently provided a publicly available photon PDF with uncertainties 
determined by those on the structure functions used as input, i.e. at the level of a few percent. 
Additionally, QED DGLAP splitting kernels have now been calculated to $\mathcal{O}(\alpha\alpha_S)$ \cite{alphaalphas} 
and $\mathcal{O}(\alpha^2)$ \cite{alpha2}. These are implicit in the LUXqed approach, and are easily implemented 
in DGLAP evolution codes.

Hence, it is now possible to be far more precise and confident about the effects of QED modified partons, the photon 
distribution and their impact on cross section calculations. The first global PDF set 
including a photon distribution based on the LUXqed approach was produced by the NNPDF group~\cite{Bertone:2017bme}. 
This was soon followed by QED corrected PDFs based on the MMHT14 PDFs (in practice also including the final HERA 
combined cross section data, so more similar to the slightly modified PDFs in \cite{MMHT2015}), which followed a 
LUXqed-inspired approach~\cite{MMHTQED} that we will summarise in the following section. More recently the CT group has also produced PDFs 
with QED corrections and a LUXqed-inspired photon distribution~\cite{Xie:2021equ}. The photon distributions in these sets 
now all have uncertainties of a few percent, and are all broadly consistent with each other, representing a huge improvement in 
the knowledge of the photon content of the proton. On the other hand, some care must be taken when claiming equivalently high precision in the corresponding PI cross sections, which in the above studies are only calculated at LO in $\alpha$; these will therefore have significantly larger  scale variation uncertainties than the percent level uncertainty due to the photon PDF. As we will discuss in the following section, of the processes entering global PDF fits, the PI contributions to off--peak lepton pair production are by far the most dominant ones. With this in mind, when calculating the PI corrections in this case we will follow the approach of~\cite{Harland-Lang:2016lhw,Harland-Lang:2021zvr}, which applies the structure function (SF) approach to directly calculate the dominant PI contribution to lepton pair production away from the $Z$ peak. This provides percent level precision in the cross section prediction here, bypassing the issue of large LO scale variations. For other processes, a standard EW K--factor approach can be taken, although in the majority of cases the impact of PI production is found to be marginal at the current level of precision.

The outline of the paper is as follows. In Section~\ref{sec:PDFmod} our procedure for including the QED effects within the MSHT framework is briefly 
described; here we start with the latest MSHT20~\cite{Bailey:2020ooq} QCD--only partons. The approach is very similar to that adopted in \cite{MMHTQED}, but involves a minor change to the
photon splitting function, the inclusion of more PI initiated contributions to fitted data sets, and the application of the SF approach to calculate these. In Section~\ref{sec:incPI} we assess the general expectations for the impact of photon--initiated (PI) corrections to processes entering  PDF fits such as MSHT20. In Section~\ref{sec:fitquality} we present the fit quality of the QED corrected global fit. As before it is found that QED corrections cause a slight deterioration in 
fit quality, and we discuss this. In Section~\ref{sec:PDFs} we describe the impact on the QCD partons as well as presenting the corresponding photon PDF. We compare to the PDFs without the inclusion of QED effects, as well as to the MMHT15qed set and the results of other groups. We also examine the phenomenological consequences of QED corrections by comparing a number of benchmark cross sections to those obtained using the QCD-only PDFs. This leads to small, but 
not insignificant changes.
In Section~\ref{sec:elinel_neutron} we present the differences between the neutron PDF up and down quarks, demonstrating the effects of QED-induced isospin violation, and 
between the proton and neutron photon PDFs. In Section~\ref{sec:avail} we describe the availability of the PDFs. As well as the conventional set of QED altered PDFs, we provide grids for the photon PDF separated into its elastic and inelastic components, and a consistent set of QED corrected neutron PDFs. Finally, in Section~\ref{sec:conc}, we conclude.

\section{Including QED effects in the MSHT framework}\label{sec:theory}

\subsection{Modifications to PDFs}\label{sec:PDFmod}

The treatment of QED corrections follows that of the MMHT15qed set, outlined in~\cite{MMHTQED}. In particular, the photon PDF is calculated using a suitable modification of the LUXqed formula~\cite{LUXQED1,LUXQED2}:
 \begin{equation}\label{eq_input}
        \begin{split}
            x\gamma(x,Q_0^2) = \frac{1}{2\pi\alpha(Q_0^2)}\int_x^1\frac{dz}{z}\Big\{ \int_{\frac{x^2 m_p^2}{1-z}}^{Q_0^2}\frac{dQ^2}{Q^2}\alpha^2(Q^2)\bigg[\bigg(zP_{\gamma,q}(z)+\frac{2x^2m_p^2}{Q^2}\bigg)F_2(x/z,Q^2)\\-z^2 F_L(x/z,Q^2)\bigg]-\alpha^2(Q_0^2)\bigg(z^2+\ln(1-z) zP_{\gamma,q}(z)-\frac{2x^2m_p^2z}{Q_0^2}\bigg)F_2(x/z,Q_0^2)\Big\}\;,
        \end{split}
    \end{equation}
    at input scale $Q_0=1$ GeV. The structure functions $F_{2,L}$ receive contributions from both elastic and inelastic photon emission, and are precisely determined using experimental data on lepton--proton scattering. In more detail, the elastic structure functions are determined from a fit by the A1 collaboration~\cite{A1:2013fsc}, and the inelastic via the HERMES GD11--P fit~\cite{HERMES:2011yno} and data from the CLAS collaboration~\cite{CLAS:2003iiq}, in the continuum and resonance regions, respectively. Uncertainties on the photon are included due to several sources (see~\cite{MMHTQED}) for further details):  the experimental uncertainty on the elastic structure functions, the value of the $R$ ratio used to determine the inelastic $F_L$, the value of the threshold $W$ between which the HERMES and CLAS data are used for the inelastic structure functions, the experimental uncertainty on the CLAS data for the resonance region, the uncertainty in the HERMES fit for the continuum region, and the modelling of renormalon corrections to the quark evolution. Each of these is included as a separate error eigenvector pair, while in addition the standard PDF uncertainties due to the fit of the QCD partons as in the MSHT20 set~\cite{Bailey:2020ooq} are included; these give 32 eigenvectors, and hence in total there are 38 eigenvectors for the MSHT20qed set\footnote{We in fact find that a more stable set of eigenvectors is found by fixing the 5th rather than the 6th Chebyshev polynomial associated with the down valence, but otherwise the same parameters are fixed as in the previous MSHT20 PDFs when generating the eigenvectors.}.
    
    QED corrections to  DGLAP evolution are again included up $O(\alpha^2)$ and $O(\alpha \alpha_S)$. The treatment of these is therefore broadly the same as in MMHT15qed, with however one exception. Namely, we now choose to include leptonic loop contributions to the photon--photon splitting function, which at $O(\alpha)$ is proportional to the sum
    \be
    \sum_i e_i^2 = N_C \sum_q^{n_F}e_q^2 + \sum_l^{n_L}e_l^2\;.
    \ee        
    In the MMHT15qed fit, the second term was omitted, as the inclusion of this strictly implies that we must include lepton PDFs, which in principle enter due to splittings of the form $\gamma \rightarrow l\bar{l}$. As these are not present in the MSHT framework, the inclusion of lepton loops in $P_{\gamma\gamma}$ would in particular lead to some small amount of violation of the momentum sum rule, due to absence of $l \to l \gamma$ splitting contributions.\footnote{In fact the MMHT15qed PDFs had a momentum violation of $+0.00008\%$ due to contributions to the photon from inelastic and higher twist sources above
$Q_0^2$. The effect from lepton splitting from photons is in the opposite direction, and of comparable, though slightly smaller size, so in total momentum is more closely conserved in MSHT20qed than in MMHT15qed.}  
However, as discussed in~\cite{MMHTQED} (see Fig.~24 of that paper) the impact of such lepton loops is not completely negligible on the photon PDF itself, reducing it by up to $2\%$ at LHC scales; the $P_{\gamma\gamma}$ splitting function leads to a reduction in the photon as it undergoes DGLAP evolution, and leptonic loop contributions increase this effect. Indeed the lack of these contributions, which are included in both the NNPDF31luxqed~\cite{Bertone:2017bme}, and CT18qed, CT18lux~\cite{Xie:2021equ} sets, can be seen in Fig.~10  of~\cite{Xie:2021equ}  to potentially lead to a rather systematic offset of the MMHT15qed photon PDF with respect to these sets. Moreover, as discussed in~\cite{Harland-Lang:2016lhw} the factorization scale of the photon PDF is directly tied to the scale at which one should evaluate the renormalization scale of $\alpha$ associated with the coupling of the photon to the production subprocess. Given the running of $\alpha$ will of course in general include lepton loops, their absence in $P_{\gamma\gamma}$ is potentially inconsistent. For these reasons, we now choose to include them. We will comment on their impact in the following sections.

\subsection{Including photon--initiated production in a global PDF fit}\label{sec:incPI}

In addition to modifying the DGLAP evolution of the QCD partons, QED corrections will also enter directly into the calculation of the  cross section predictions entering the fit. This will include  PI production, which  forms a subset of the broader class of EW corrections. It is therefore useful to assess on general grounds the size that such PI contributions are expected to have for processes that enter current global PDF fits. 

\begin{figure} 
\begin{center}
\includegraphics[scale=0.68]{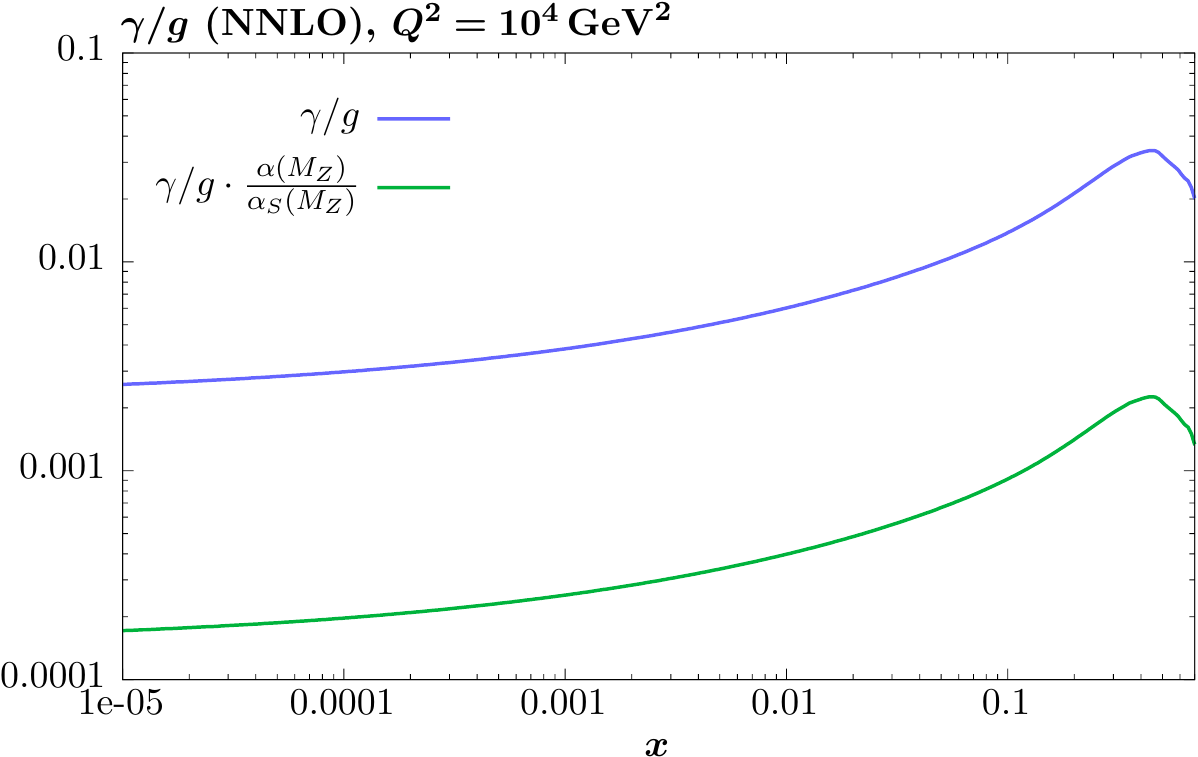}
\includegraphics[scale=0.68]{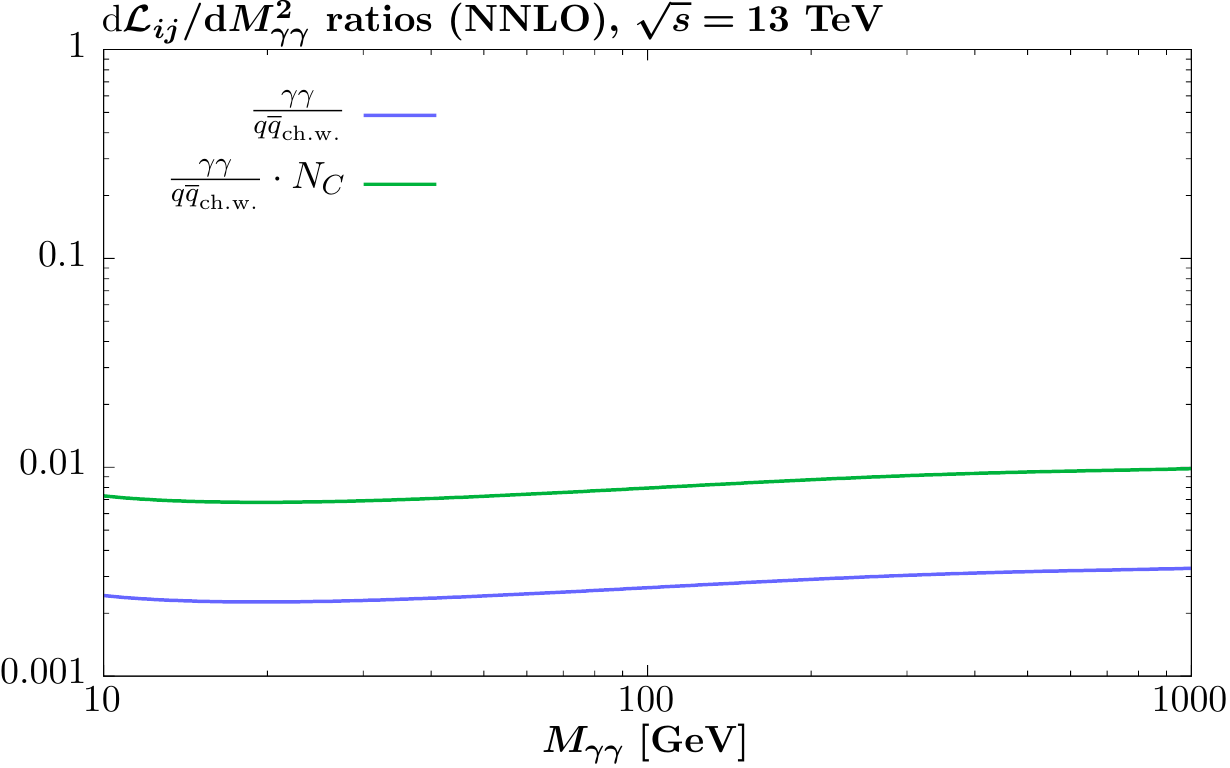}
\caption{\sf (Left) The ratio of the $\gamma$ to the $g$  distributions at $Q^2=10^4$ ${\rm GeV}^2$. In addition the same ratio, but weighted by $\alpha(M_Z)/\alpha_S(M_Z)$, is shown. (Right) Ratio of the $\gamma\gamma$ to the charge weighted $q\overline{q}$ luminosity at 13 TeV. Also shown is the same ratio, weighted by inverse of the LO QCD DY colour factor. In both plots the PDFs result from NNLO fits to the MSHT20 dataset, with QED effects included}
\label{fig:q_vs_g}
\end{center}
\end{figure}

As discussed above, the PI channel can play an important role in the production of objects with EW couplings, such as EW boson--boson scattering and Higgs boson production with an associated EW boson; however, these are not currently included in global PDF fits. Of the processes entering the global fits, such as MSHT20, one might broadly consider DIS, inclusive jets, $t\overline{t}$ and lepton pair production as receiving potentially relevant PI contributions. For the production of strongly interacting particles in hadron--hadron collisions such as $t\overline{t}$ and jets, the leading PI contribution comes from replacing an initial--state gluon with a photon, for all relevant diagrams, e.g. for $gg \to t\overline{t}$ we instead have $\gamma g \to t\overline{t}$. To estimate the expected suppression in this case in Fig.~\ref{fig:q_vs_g} we show the ratio of the photon to gluon PDFs at a representative scale $Q^2 = 10^4$ ${\rm GeV}^2$ (although the precise choice does not effect the results significantly). We can see that photon PDF is suppressed by $\sim 2$ orders of magnitude or more, with the exception of the largest $x$ values where the suppression  is a little less, though still significant. This is in particular a significantly larger suppression than one might rather naively expect from simple $\sim \alpha/\alpha_S$ scaling in the corresponding PDFs. In the $g \gamma$--initiated process, we will receive  an additional suppression  $\sim \alpha/\alpha_S$ due to the $\gamma q$ coupling, and hence we also show the same ratio, but weighted by this factor. We can see that the expected correction enters at the per mille level, and hence in practice may be expected to enter at roughly the same level as ${\rm N}^3$LO QCD corrections or even less, rather than the NNLO order one might naively expect from the fact that $\alpha_S(M_Z)^2 \sim \alpha(M_Z)$.

Now in reality, the true result will of course require a full calculation, accounting for the colour factors, differing diagrams that enter and the fractional quark charge; the former effects will broadly speaking enhance the PI contribution, whereas the latter will suppress it. This therefore should, and can, be verified by explicit calculation. This is achieved  in e.g.~\cite{Bertone:2017bme} for the case of $t\overline{t}$ production, and in fact the level of suppression is found to be larger than that expected from Fig.~\ref{fig:q_vs_g} (left), by roughly a further order of magnitude. For jet production, we can expect the suppression to be even greater, due to the presence of pure gluonic channels, which of course receive no leading PI contribution, although the heavy top quark mass in the $t\overline{t}$ case makes the comparison a little less direct. For DIS, the same argument applies as above, but now it is the NLO in QCD gluon--initiated diagram that receives the leading correction as above, and hence the suppression is expected to be an $O(\alpha_S)$ more. We therefore conclude that the impact of PI contributions to these processes is in general not expected to be significant at the NNLO QCD precision level\footnote{A similar argument can be applied to the case of e.g. isolated photon production, which is included in the NNPDF4.0 fit~\cite{Ball:2021leu}.}, although this should of course be verified by explicit calculation, given it may not apply uniformly in all kinematic regions and for all processes.

We are therefore left to consider lepton pair production, for which the final state is of course not strongly interacting, and hence the arguments above do not apply. Here, as is well known, the $t$--channel $\gamma\gamma\to l^+ l^-$ process can be of much greater phenomenological relevance. To demonstrate this, in Fig.~\ref{fig:q_vs_g} (right) we show the ratio of the $\gamma\gamma$ partonic luminosity to the $e_q^2$ charge weighted $q\overline{q}$ luminosity, relevant for $q\overline{q} \to \gamma^*$ DY production; see e.g.~\cite{Campbell:2006wx} for a definition of these. We can see that this ratio is at the level of a few per mille, however once we divide by the LO QCD DY colour factor, $1/N_C$, this is at the level of 1\%, as can be seen in the figure. When we also account for the $t$--channel enhancement of the $\gamma\gamma$ cross section:
\be
\frac{|\mathcal{M}(\gamma\gamma \to l^+ l^-)|^2}{|\mathcal{M}(q\overline{q} \to l^+ l^-)|^2} \propto \frac{\hat{s}^2}{\hat{u}\hat{t}} \geq 4 \;,
\ee
where $\hat{s},\hat{t},\hat{u}$ are the usual partonic Mandlestam variables, we can expect this to enter at the $\sim 5-10\%$ level, which indeed it is seen to~\cite{Bertone:2017bme,Harland-Lang:2019eai,Harland-Lang:2021zvr}. On the other hand, in the $Z$ peak region the DY process receives a significant resonant enhancement, and hence the above conclusions do not hold; as demonstrated in~\cite{Bertone:2017bme,Harland-Lang:2019eai,Harland-Lang:2021zvr} in this region the relative contribution from the $\gamma\gamma \to l^+ l^-$ subprocess is at the per mille level. We note that there are in addition $\gamma \to q \overline{q}$ splittings that can play a role here, that is from the mixed $\gamma q(\overline{q})$ initial state. However, here we are are once again in the situation described above for case of $t\overline{t}$ and jet production, and so again will expect per mille level corrections from this; in fact as the corresponding $g \to  q \overline{q}$ diagram now only enters at NLO in QCD we might expect the suppression to be larger still, although this will depend on e.g. the impact of colour factors and fractional quark charges. In e.g.~\cite{Arbuzov:2007kp} results are presented for this channel using the MRST2004QED~\cite{mrstqed} set, and are indeed found to enter at the level of a few per mille. While this set is certainly now outdated, as seen in~\cite{LUXQED2} it lies within $\sim 50\%$ of the LUXqed photon PDF, and hence the true result will also be of this order.
As discussed further below, an additional point to note is that, for the current highest precision on--peak $Z$ data entering the fit, from the ATLAS collaboration, the $t$--channel $\gamma\gamma \to l^+ l^-$ PI contribution is in any case subtracted directly at the data level. We finally note that although the inclusive lepton pair signal has been discussed above, very similar conclusions will hold for measurements of the lepton pair $p_\perp$ distribution, or production in association with jets, which  also enter PDF fits.

We  therefore conclude that, of the processes that currently enter  global PDF fits, lepton pair production away from the $Z$ peak region receives by far the largest PI contribution, which as discussed above can enter at the 5-10\% level.  With this in mind, to calculate this we use the \texttt{SFGen} Monte Carlo implementation of the Structure Function (SF) approach discussed in~\cite{Harland-Lang:2019eai,Harland-Lang:2021zvr}. This automatically provides a percent--level precision calculation of the lepton pair production cross section in this region, without the significant scale variation uncertainties that are present in the LO collinear calculation. The underlying experimental inputs for the proton structure functions are precisely the same as in the current study for the photon PDF, and the dominant theoretical uncertainties are  due to these. In practice the SF prediction will depend on the underlying QCD parton set used to calculate the inelastic structure functions in the high $Q^2$ region (as is the case for the photon PDF), and so an iterative procedure should be adopted in the fit, where the PI cross section is recalculated using the QCD partons from each iteration of the fit, until convergence is reached. We have performed such a procedure here, although in reality we find this converges after one iteration, and we have confirmed that calculating the PI cross section with the MSHT20 QCD partons, without such an iteration, gives an almost identical result in terms of the fit quality and the extracted PDF set. However, this may not remain true in future fits, where more PI corrections may be included in phenomenologically relevant regions.

In addition to the ATLAS high mass DY data, we now also include the PI contribution to the CMS 7 TeV double differential DY data~\cite{CMS-ddDY}\footnote{We recall~\cite{Bailey:2020ooq} that the 8 TeV measurement is not included in the fit, due to apparent issues in the data~\cite{CMSDDDY8}.}, which extends above and below the $Z$ peak region. There are also  ATLAS measurements of $W,Z$ production at 7 and 8 TeV~\cite{ATLASWZ7f,ATLAS8Z3D}, which extend below and above the $Z$ peak region, as well as the 8 TeV measurement of the $Z$ boson (or more precisely, dilepton) $p_\perp$ distribution. In these cases we could in principle include the PI contribution via the SF approach, however as discussed in~\cite{Bailey:2020ooq}, for these ATLAS measurements this contribution is already subtracted from the data. We note that the procedure for doing this is theory--dependent and indeed in some cases uses rather outdated photon PDF sets/calculations. It is therefore certainly recommended that such subtractions are not made for future datasets. However, as a result of this, we do not include PI contributions for these processes here. We note that in~\cite{Bailey:2020ooq} we did include the PI contribution to the ATLAS 8 TeV DY data~\cite{ATLAS8Z3D}, but as discussed above this should not have been done, given the data is subtracted. This has now been removed, and indeed doing so improves the fit quality, as we will see. However the impact on the PDFs is generally very small, and hence while for completeness in Section~\ref{sec:PDFs} we will compare against a QCD--only set resulting from a fit with this correction applied, in reality for public use one can take the MSHT20nnlo set as a baseline for comparison, and indeed for wider use; we therefore do not make this QCD--only set available for release.

Finally, we note that one could alternatively include a calculation of the PI contribution to lepton pair production in collinear factorization that goes beyond LO in $\alpha$, in order to improve the precision of the large LO scale variation uncertainty. However, as discussed in~\cite{Harland-Lang:2021zvr}, even at NLO the scale variation in this case can remain larger than the PDF uncertainty. Therefore, given the rather large size of these contributions away from the $Z$ peak region, we consider the use of the SF calculation as the more appropriate choice here. 

On the other hand, when considering the broader class of PI observables that currently enter a global PDF fit, for example lepton pair production in the $Z$ peak region (where initial--state $\gamma \to q\overline{q}$ splittings become relevant), $t\overline{t}$ or jet production, one can more straightforwardly make use of collinear factorization. Indeed, in practice this becomes essential, as here the PI contribution will in general form a subset of the larger class of QED/EW corrections. These can more broadly be relevant at the NNLO QCD precision level, even if the dominant contribution from these often comes from EW Sudakov logarithms that are therefore unrelated to pure QED corrections and PI production. The conclusion in this case is that, for the processes such as those considered above, where PI production enters at the per mille level, it is certainly appropriate to include them as part of the broader NLO EW corrections in a standard K--factor manner, provided a photon PDF based on the LUXqed approach is used to calculate these K--factors. If this is the case, then the difference due to the precise choice of photon PDF will be at the level of a few percent of a per mille correction, i.e. clearly negligible and in any case significantly less than the scale variation uncertainty in a LO PI calculation. 

In the MSHT20 fit, we include EW corrections for a range of processes. In more detail, for inclusive jet production we include these for all 7 and 8 TeV LHC datasets~\cite{ATLAS7jets,CMS7jetsfinal,CMS8jets} using K--factors evaluated from the calculation of~\cite{Dittmaier:2012kx}. These do not include QED corrections, and therefore PI production, arguing that the dominant contribution is from the pure weak corrections (a distinction that can be made in a gauge invariant way in this case), due to their Sudakov logarithmic enhancement; the size of the overall EW corrections, which is driven by this, can be as large as $\sim 10\%$ at the highest jet $p_\perp$ values, though it is generally rather  less. 

For the ATLAS 8 TeV $Z$ $p_\perp$ data we apply the same EW K--factors as used in the ATLAS analysis~\cite{ATLASZpT}, which are derived from the calculation of~\cite{Denner:2011vu}. These include mixed $\gamma q$ PI production, and  are found to enter at the per mille level and be significantly smaller than the other EW corrections. However, these make use of the now outdated MRST2004QED~\cite{mrstqed} set, and hence in principle it should not be relied upon. In practice, as discussed above this set lies within $\sim 50\%$ of the LUXqed photon PDF. Hence any update to account for this would only modify the eventual EW K--factor at the per mille level, and possibly less. We therefore for simplicity continue to make use of these K--factors, which correctly account for the dominant EW correction. In future analyses a photon PDF based on the LUXqed formalism should be used to calculate such K--factors, although in reality  as discussed above the precise choice will not matter. The total size of EW corrections is as large as $\sim 20\%$ at high $p_\perp^{ll}$, though is generally less than this~\cite{ATLASZpT}.

For the ATLAS high precision $W, Z$ data~\cite{ATLASWZ7f} we apply the same EW K--factors as used in the ATLAS analysis. These also include mixed $\gamma q$ corrections, in this case derived from the \texttt{MCSANC} generator~\cite{Bondarenko:2013nu,Arbuzov:2015yja}, but again using the MRST2004QED set. However, once again these are observed to enter at the per mille level, and hence we can safely apply these, even if future calculations should use an updated photon PDF set. The total size of the EW corrections is $\sim 0.5\%$ at intermediate and high masses, but $\sim 6\%$ in the lowest mass region~\cite{ATLASWZ7f}.

Finally, for the ATLAS and CMS single differential top quark pair production data~\cite{ATLASsdtop,CMSttbar08_ytt} we use the EW K--factors calculated in~\cite{NNLOtopEW}. These include the $\gamma g$ initiated channel, calculated using the LUXqed photon PDF~\cite{LUXQED1}. However, entirely consistently with the discussion above, this contribution is found to be negligible. 
\section{Fit Quality}\label{sec:fitquality}

\begin{table}
\begin{center}
\begin{tabular}{l|c|c|c|}\hline \hline
  Data set & QCD & QED & Change \\ \hline
  BCDMS $\mu p$ $F_2$ \cite{BCDMS} &  178.8/163 & 182.5/163 & \textcolor{red}{(+3.7)}  \\ 
  BCDMS $\mu d$ $F_2$ \cite{BCDMS} &  145.6/151 & 146.8/151 & \textcolor{red}{(+1.2)}  \\ 
  NMC $\mu p$ $F_2$ \cite{NMC} &  124.0/123 & 124.9/123 & - \\ 
  NMC $\mu d$ $F_2$ \cite{NMC} &  112.4/123 & 113.1/123 & - \\ 
  NMC $\mu n/\mu p$ \cite{NMCn/p} & 130.4/148 & 128.9/148  & \textcolor{blue}{(-1.5)}  \\ 
  E665 $\mu p$ $F_2$ \cite{E665} & 65.0/53 & 65.0/53 & - \\ 
  E665 $\mu d$ $F_2$ \cite{E665} &  59.9/53 & 59.7/53 & -\\   
  SLAC $e p$ $F_2$ \cite{SLAC,SLAC1990} &  32.3/37 & 32.4/37 &  -\\   
  SLAC $e d$ $F_2$ \cite{SLAC,SLAC1990} &  22.9/38 & 23.0/38 &-\\     
  NMC/BCDMS/SLAC/HERA $F_L$ \cite{NMC,BCDMS,SLAC1990,H1FL,H1-FL,ZEUS-FL}  & 68.4/57 & 68.2/57 &- \\ \hline
  E866/NuSea $pp$ DY \cite{E866DY} &  225.8/184 & 226.0/184 & - \\
  E866/NuSea $pd/pp$ DY \cite{E866DYrat} &  9.5/15 & 8.8/15 & -\\ \hline
  NuTeV $\nu N$ $F_2$ \cite{NuTeV}  & 38.2/53 & 37.2/53 & \textcolor{blue}{(-1.0)} \\
  CHORUS $\nu N$ $F_2$ \cite{CHORUS}  & 30.3/42 & 29.4/42 & - \\
  NuTeV $\nu N$ $x F_3$ \cite{NuTeV}  & 30.9/42 & 30.5/42 &\\  
  CHORUS $\nu N$ $x F_3$ \cite{CHORUS}  & 18.4/28 & 18.4/28 &- \\
  CCFR $\nu N \rightarrow \mu \mu X$ \cite{Dimuon} &  68.1/86 & 68.4/86 &- \\
  NuTeV $\nu N \rightarrow \mu \mu X$ \cite{Dimuon} &  57.5/84 & 56.7/84 &\textcolor{blue}{(-1.0)} \\ \hline
  HERA $e^+ p$ CC \cite{H1+ZEUS} &  50.2/39 & 50.9/39 &- \\
  HERA $e^- p$ CC \cite{H1+ZEUS} &  70.3/42 & 72.2/42 &\textcolor{red}{(+1.9)} \\
  HERA $e^+ p$ ${\rm NC}~820$~GeV \cite{H1+ZEUS} & 89.9/75 & 90.1/75 &- \\ 
  HERA $e^+ p$ ${\rm NC}~920$~GeV \cite{H1+ZEUS} & 510.7/402 & 511.2/402 &-  \\
  HERA $e^- p$ ${\rm NC}~460$~GeV \cite{H1+ZEUS} & 247.6/209 & 248.0/209 &- \\
  HERA $e^- p$ ${\rm NC}~575$~GeV \cite{H1+ZEUS} & 262.2/259 & 262.8/259 &- \\
  HERA $e^- p$ ${\rm NC}~920$~GeV \cite{H1+ZEUS} & 243.9/159 & 244.8/159 &- \\  
  HERA $e p$ $F_{2}^{\text{charm}}$ \cite{HERAhf} & 132.6/79 & 131.9/79 &- \\  
  D{\O} II $p\bar{p}$ incl. jets \cite{D0jet} &  120.3/110 & 119.5/110 &- \\
  CDF II $p\bar{p}$ incl. jets \cite{CDFjet} &  60.1/76 & 60.9/76 &- \\
  CDF II $W$ asym. \cite{CDF-Wasym} &  18.9/13 & 18.5/13 & -\\
  D{\O} II $W\rightarrow \nu e$ asym. \cite{D0Wnue} &  33.5/12 & 33.5/12 &- \\
  D{\O} II $W \rightarrow \nu \mu$ asym. \cite{D0Wnumu} &  17.8/10 & 17.6/10 & -\\
  D{\O} II $Z$ rap. \cite{D0Zrap}   & 16.3/28 & 16.4/28 & -\\
  CDF II $Z$ rap. \cite{CDFZrap}  & 37.1/28 & 37.2/28 &- \\
  D{\O} $W$ asym. \cite{D0Wasym}  & 12.8/14 & 11.3/14  &\textcolor{blue}{(-1.5)} \\ \hline
 \hline
\end{tabular}
\end{center}
\caption{\sf The values of $\chi^2/N_{\rm pts}$ for the non-LHC data sets included in a NNLO fit to the MSHT20 dataset, with and without QED corrections. The difference in $\chi^2/N$pts is also shown explicitly, for the cases that the magnitude is larger than 1 point; negative values indicate a better fit quality in the QED case.}
\label{tab:chisqtable}
\end{table}

\begin{table} 
\begin{center}
\begin{tabular}{l|c|c|c|}\hline \hline
  Data set & QCD & QED & Change \\ \hline
  ATLAS $W^+$, $W^-$, $Z$ \cite{ATLASWZ} &  29.9/30 &  29.7/30 & -\\ 
  CMS $W$ asym. $p_T > 35$~GeV \cite{CMS-easym} &  8.2/11 &  8.0/11 & -\\ 
  CMS asym. $p_T > 25, 30$~GeV \cite{CMS-Wasymm} &  7.4/24 &  7.5/24 &- \\ 
  LHCb $Z\rightarrow e^+e^-$ \cite{LHCb-Zee} &  22.3/9 &  22.6/9 &- \\ 
  LHCb $W$ asym. $p_T > 20$~GeV \cite{LHCb-WZ} &  12.4/10 &  12.1/10 & -\\ 
  CMS $Z\rightarrow e^+e^-$ \cite{CMS-Zee} &  18.0/35 &  18.0/35 &- \\ 
  ATLAS High-mass Drell-Yan \cite{ATLAShighmass} &  18.6/13 &  19.1/13 &- \\   
  CMS double diff. Drell-Yan \cite{CMS-ddDY} &  144.8/132 &  145.4/132 &- \\   
  Tevatron, ATLAS, CMS $\sigma_{t\bar{t}}$ \cite{Tevatron-top}-\cite{CMS-top8} &  14.5/17 &  14.4/17 &- \\     
  LHCb 2015 $W$, $Z$ \cite{LHCbZ7,LHCbWZ8} &  101.4/67 & 100.6/67 &- \\
  LHCb 8~TeV $Z\rightarrow ee$ \cite{LHCbZ8}  & 26.2/17 & 25.8/17 &- \\
  CMS 8~TeV $W$ \cite{CMSW8} &  12.6/22 & 14.0/22 &\textcolor{red}{(+1.4)} \\
  ATLAS 7~TeV jets \cite{ATLAS7jets} & 221.3/140 & 217.8/140 & \textcolor{blue}{(-3.5)}\\
  CMS 7~TeV $W+c$ \cite{CMS7Wpc} & 8.3/10 & 7.8/10 &- \\
  ATLAS 7~TeV high precision $W$, $Z$ \cite{ATLASWZ7f} & 117.3/61 & 119.4/61 &\textcolor{red}{(+2.1)} \\
  CMS 7~TeV jets \cite{CMS7jetsfinal} &  176.9/158 & 176.4/158 &- \\
  CMS 8~TeV jets \cite{CMS8jets} &  262.4/174 & 267.9/174 &\textcolor{red}{(+5.5)} \\
  CMS 2.76~TeV jet \cite{CMS276jets} &  102.4/81 & 102.8/81 &- \\
  ATLAS 8~TeV $Z$ $p_T$ \cite{ATLASZpT} &  190.8/104 & 200.8/104 &\textcolor{red}{(+10.0)} \\
  ATLAS 8~TeV single diff $t\bar{t}$ \cite{ATLASsdtop} &  25.8/25 & 26.8/25 &\textcolor{red}{(+1.0)}  \\
  ATLAS 8~TeV single diff $t\bar{t}$ dilepton \cite{ATLASttbarDilep08_ytt} &  3.3/5 & 3.7/5 &- \\
  CMS 8~TeV double differential $t\bar{t}$ \cite{CMS8ttDD} &  22.3/15 & 22.1/15 &- \\
  CMS 8~TeV single differential $t\bar{t}$ \cite{CMSttbar08_ytt} &  13.0/9 & 13.3 /9 &-  \\
  ATLAS 8~TeV High-mass Drell-Yan \cite{ATLASHMDY8} &56.9/48 & 57.8/48 &-  \\
  ATLAS 8~TeV $$W$$ \cite{ATLASW8} &  57.8/22 & 59.2/22 &\textcolor{red}{(+1.4)}  \\
  ATLAS 8~TeV $W+\text{jets}$ \cite{ATLASWjet}  & 18.7/30 & 19.1/30 & -\\
  ATLAS 8~TeV double differential $Z$ \cite{ATLAS8Z3D} & 75.9/59 & 77.5/59 &\textcolor{red}{(+1.6)}  \\ \hline
  Total & 5111.8/4363 & 5136.1/4363  &\textcolor{red}{(+24.3)}   \\ 
 \hline
\end{tabular}
\end{center}
\caption{\sf The values of $\chi^2/N_{\rm pts}$ for the LHC data sets included in a NNLO fit to the MSHT20 dataset, with and without QED corrections. The difference in $\chi^2/N$pts is also shown explicitly, for the cases that the magnitude is larger than 1 point; negative values indicate a better fit quality in the QED case. The total $\chi^2$ value corresponds to the sum of the individual values shown in Tables~\ref{tab:chisqtable} and~\ref{tab:LHCchisqtable}.}
\label{tab:LHCchisqtable}
\end{table}

In this section we describe the changes in fit quality in the NNLO global fit from the addition of QED effects, with a fixed value of $\alpha_S(M_Z)=0.118$, as this corresponds to the default fit value. Allowing this to be free gives a marginal improvement in the fit quality, by about 1 point in $\chi^2$, and a preferred value of $\alpha_S(M_Z)=0.1176$, that is the same at the quoted level of precision to the case of the pure QCD fit. Table~\ref{tab:chisqtable} provides the comparison for the non-LHC datasets, whilst Table~\ref{tab:LHCchisqtable} shows the LHC datasets. The final column of each table highlights the main changes in the fit qualities in terms of the $\chi^2$ of each dataset. Overall there is a worsening of the fit quality upon inclusion of QED effects of approximately 24 points in $\chi^2$. In the previous  MMHT2015qed fit~\cite{MMHTQED}, the fit quality was also found to deteriorate, but by a rather smaller amount of $\sim 7$ points, albeit for rather fewer data points. Indeed, there are now substantially more datasets included in the global fit, with in particular various high precision LHC datasets being included.  Some of the changes in $\chi^2$ are similar to before, with the BCDMS and HERA $e^-p$ data fit quality deteriorating by similar amounts to the MMHT2015qed case, whilst the NuTeV $F_2$ data improves by a comparable amount to that observed in MMHT2015qed. In the BCDMS case this is due to $q \to q\gamma$ emission, which leads to a quicker high-$x$ quark evolution, i.e. mimicking a slightly larger value of $\alpha_S$,  which the BCDMS data is known to disfavour~\cite{Cridge:2021qfd}.

The new datasets in MSHT20 which show the largest sensitivity to the presence of QED effects in the fit are the ATLAS 8~TeV $Z$ $p_T$, CMS 8~TeV inclusive jets and ATLAS 7~TeV inclusive jets datasets, with the first two of these worsening by 10 and 5.5 points in $\chi^2$ and the last improving by 3.5 points. All three of these datasets are precise and sensitive to the shape of the gluon in the high $x$ region, and given this is altered by the addition of QED effects in the refit (see Fig.~\ref{fig:s_gluon} (right) later) it is perhaps unsurprising that their fit qualities alter. For the inclusive jet datasets, given the relatively small changes in the $\chi^2$ and their sensitivity to a range of $x$ values, it is difficult to be precise about the causes of the improvement in the ATLAS 7~TeV jets and the worsening of the CMS 8~TeV jets. Nonetheless, we have seen previously that these datasets pull on the high $x$ gluon in different ways and display some tension, therefore it is consistent at least that one worsens in fit quality whilst the other improves. As for the ATLAS 8~TeV $Z$ $p_T$ data, this has already been demonstrated to be in tension with many of the datasets sensitive to the high $x$ gluon in the MSHT20 fit~\cite{Bailey:2020ooq}. This tension clearly worsens upon the inclusion of the QED effects, and indeed if this dataset is removed the deterioration in fit quality when including QED effects reduces to 12 points in $\chi^2$, more similar to the effect of the inclusion of QED effects observed in MMHT2015qed. 

Further changes in fit quality tend to be small and spread out across the datasets, with several non-LHC datasets improving marginally, while many of the newer LHC datasets worsen in fit quality very slightly upon the addition of QED effects, including the ATLAS 7 and 8~TeV high precision $W,Z$ measurements. We note that the fit quality for the QCD fit to the ATLAS 8 TeV DY data~\cite{ATLAS8Z3D} is $ \chi^2 \sim 76$, which is $\sim 10$ points better than the value reported in~\cite{Bailey:2020ooq}. As discussed in the previous section, this is due to the removal of the PI contribution, which was incorrectly included in the MSHT20 fit, given this is subtracted from the data. 

We note that, as discussed in the previous section, explicit PI contributions are only included for two datasets, namely the ATLAS 8 TeV high mass DY~\cite{ATLASHMDY8} and the CMS 7 TeV double differential DY~\cite{CMS-ddDY}. The size of the PI contributions to these are as much as $5\%$ of the QCD DY prediction, depending on the mass and rapidity region, and hence it is interesting to investigate the impact this has on the fit quality, in addition to that due to QED corrections to DGLAP evolution. We have therefore repeated the fit with the the PI components for these datasets excluded, as well as with them included but using LO collinear factorization (with $\mu_F=\mu_R = m_{ll}$), as opposed to the SF approach. For the ATLAS data, we find that including the PI component via LO collinear factorization, or even excluding the PI component entirely, has a very mild impact, at the level of less than 1 point in $\chi^2$. For the CMS data, on the other hand, we find that excluding the PI component leads to a deterioration of the fit quality by $\sim 10$ points in $\chi^2$. Interestingly, a very similar level of deterioration is seen if we instead include the PI component via LO collinear factorization. Therefore, we can conclude that there is a clear preference for the SF approach here, although the impact on the final PDF fit will be very small.

\section{Impact on PDFs and Benchmark Cross Sections}\label{sec:PDFs}

\begin{figure} 
\begin{center}
\includegraphics[scale=0.65]{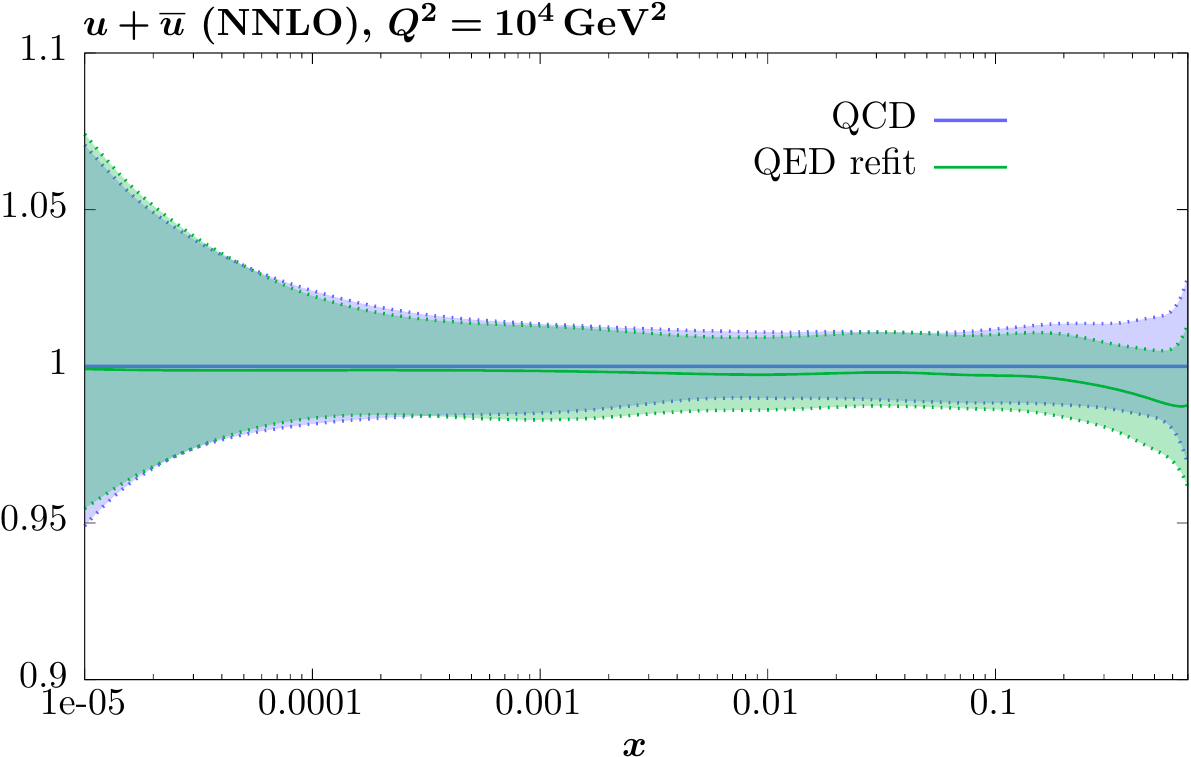}
\includegraphics[scale=0.65]{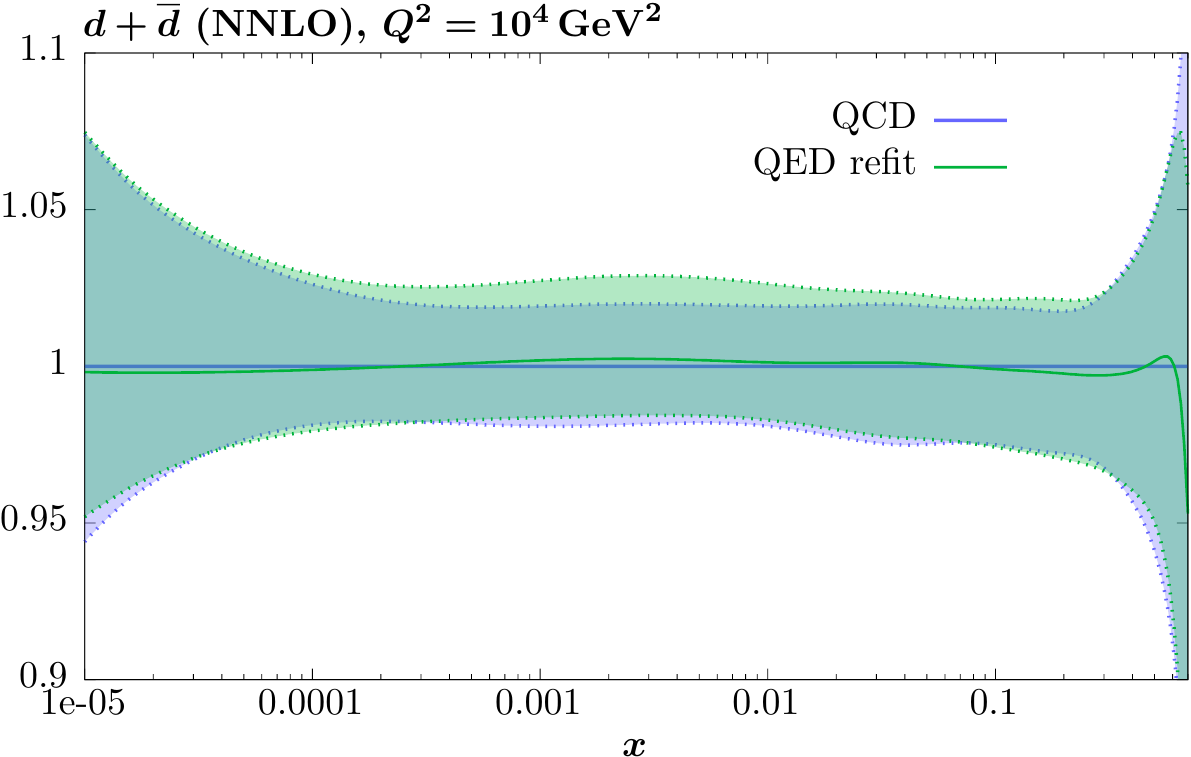}
\caption{\sf The ratio of the $u+\overline{u}$ and $d+\overline{d}$  distributions (with uncertainties) at $Q^2=10^4$ ${\rm GeV}^2$, resulting from NNLO fits to the MSHT20 dataset, with QED effects included to that without.}
\label{fig:q_p_qbar}
\end{center}
\end{figure}

\begin{figure} [t]
\begin{center}
\includegraphics[scale=0.65]{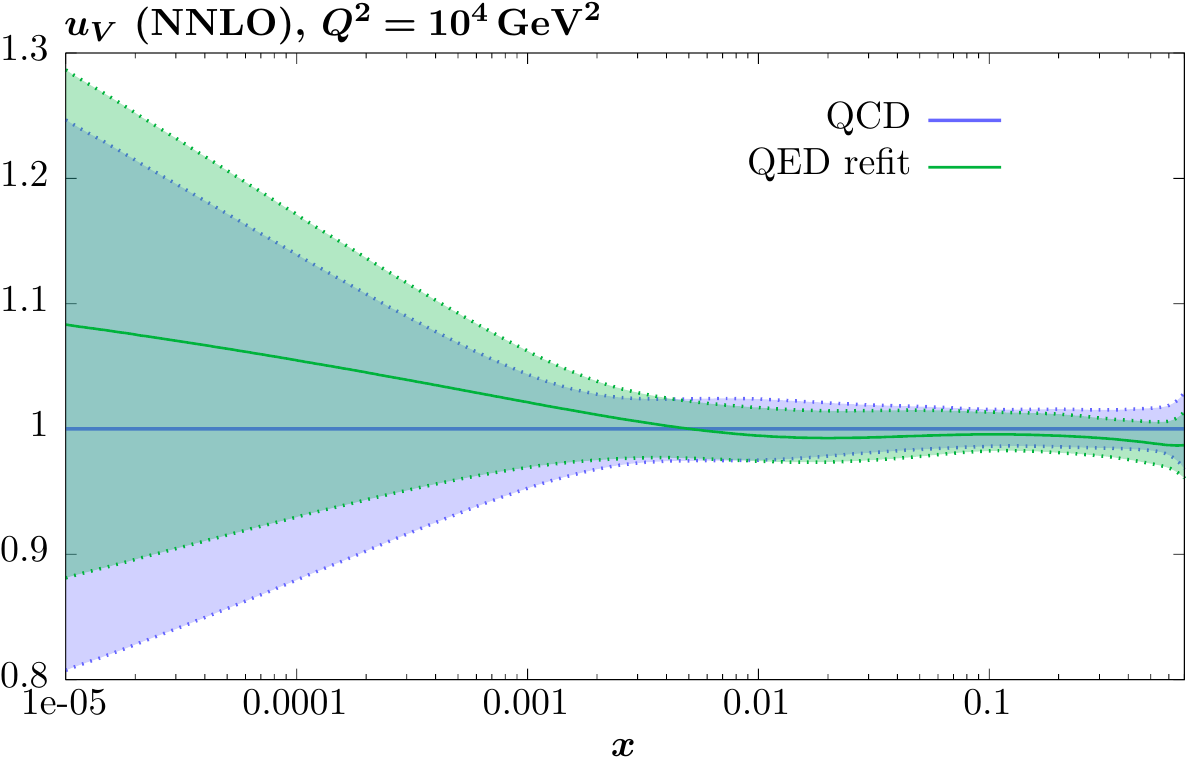}
\includegraphics[scale=0.65]{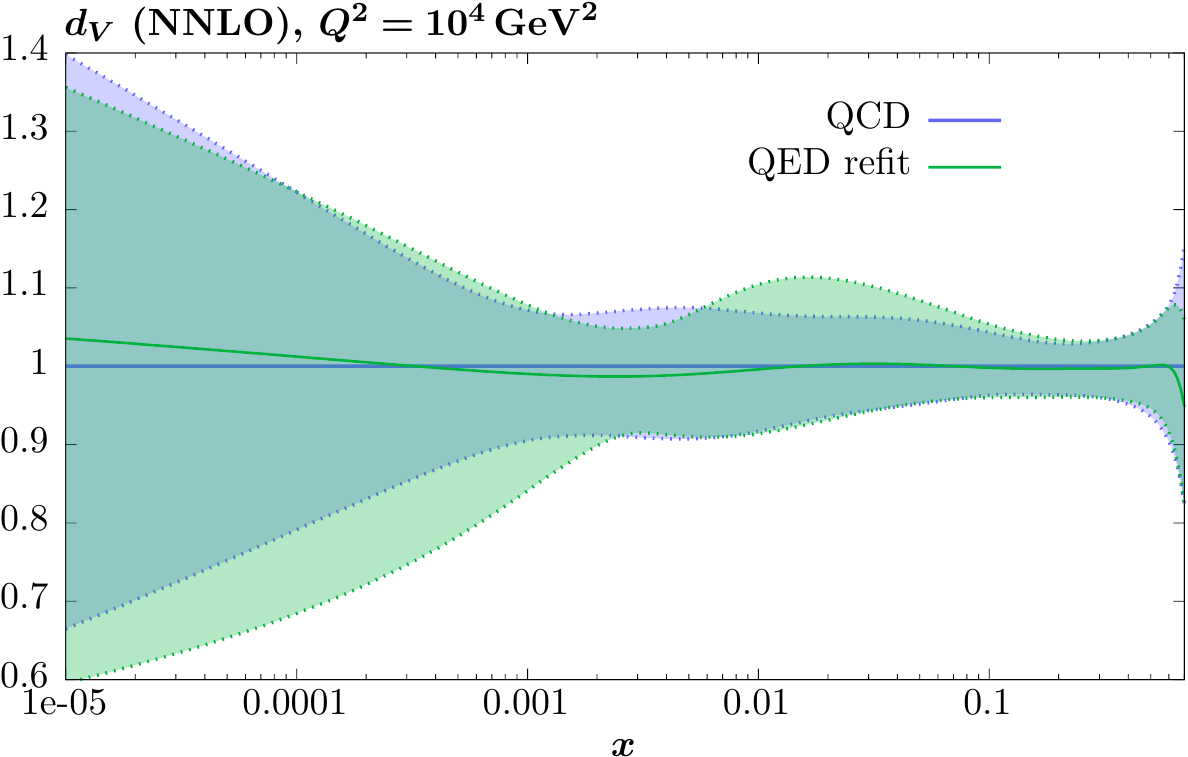}
\caption{\sf The ratio of the $u_V$ and $d_V$  distributions (with uncertainties) at $Q^2=10^4$ ${\rm GeV}^2$, resulting from NNLO fits to the MSHT20 dataset, with QED effects included to that without.}
\label{fig:qv}
\end{center}
\end{figure}

\begin{figure} [t]
\begin{center}
\includegraphics[scale=0.65]{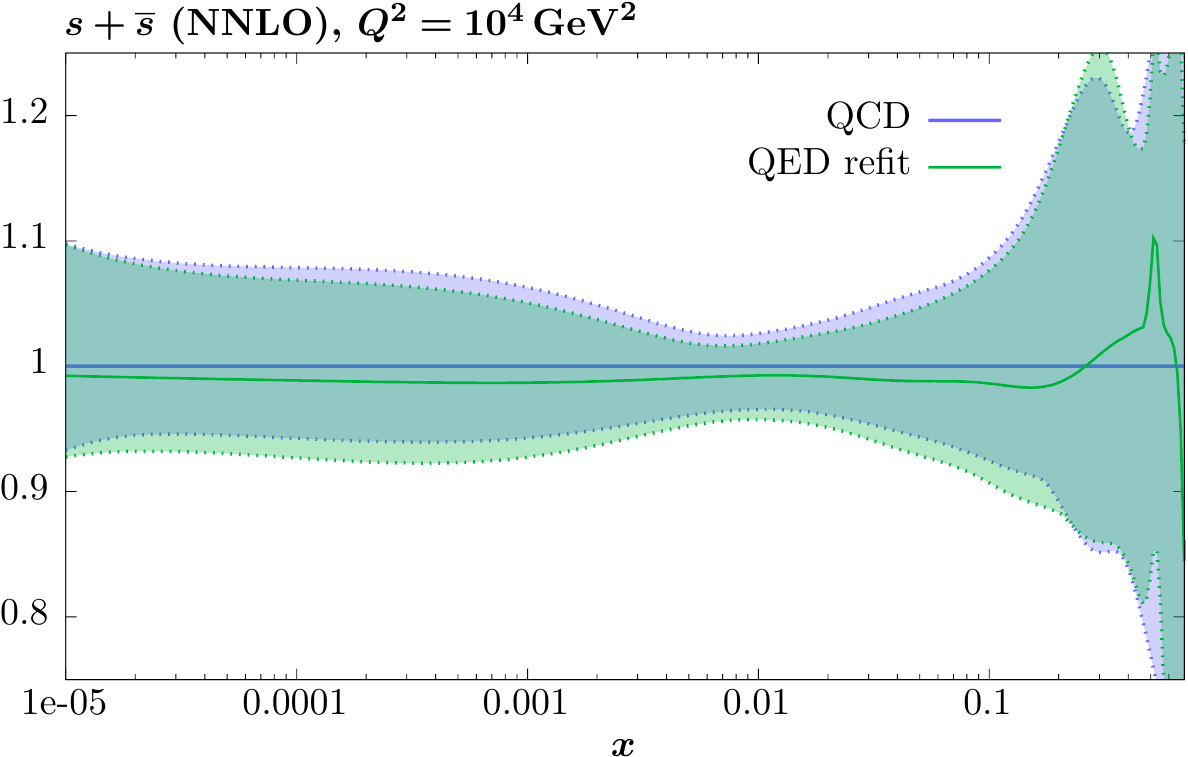}
\includegraphics[scale=0.65]{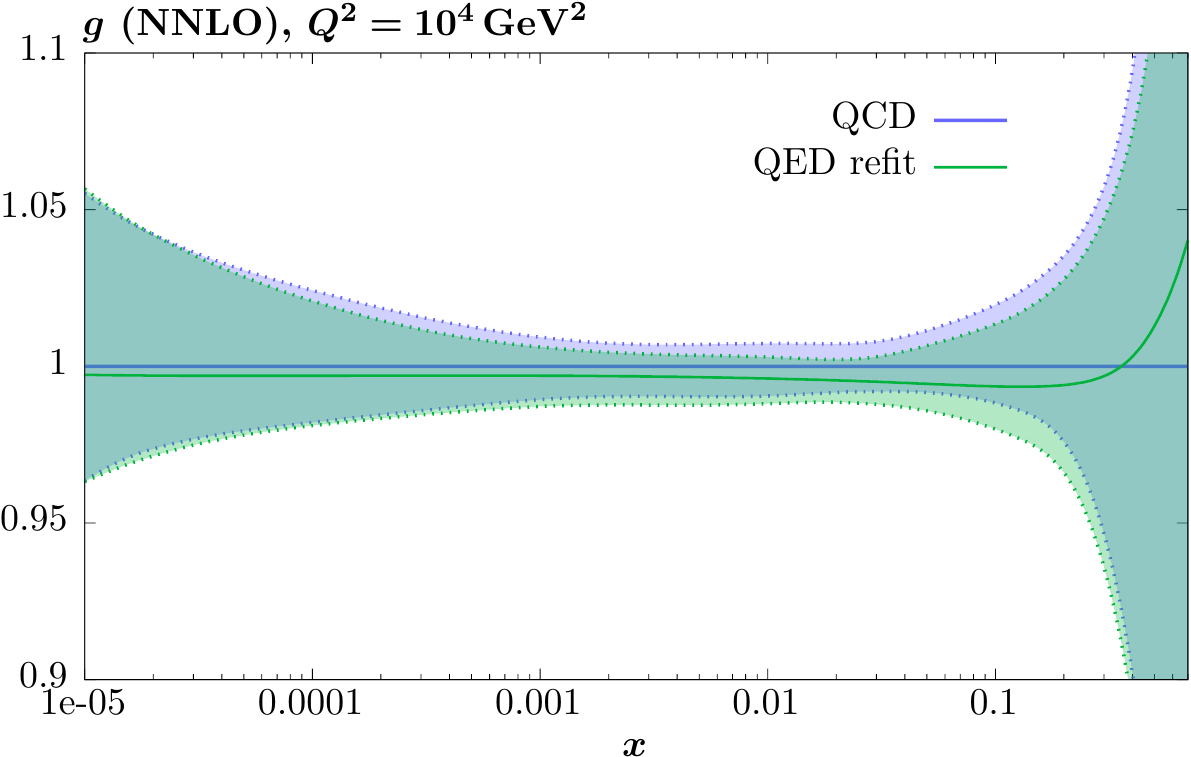}
\caption{\sf The ratio of the $s+\overline{s}$ and $g$  distributions (with uncertainties) at $Q^2=10^4$ ${\rm GeV}^2$, resulting from NNLO fits to the MSHT20 dataset, with QED effects included to that without.}
\label{fig:s_gluon}
\end{center}
\end{figure}

We now present the impact of the inclusion of QED effects on the MSHT PDFs. Many of the effects mimic those seen in the MMHT15qed set, and therefore we show only a selection here. In all cases, these correspond to the scale $Q^2=10^4$ ${\rm GeV}^2$.

We begin with the up and down singlet distributions, $u+\bar{u}$ and $d+\bar{d}$, in Fig.~\ref{fig:q_p_qbar}. At high $x$ these may be expected to show a reduction in the PDFs due to $q \rightarrow q + \gamma$ emission. This reduces the quark singlet momenta, and correspondingly increases the photon PDF, with the effect being most pronounced at high $x$. This can be clearly seen in the up singlet distribution in Fig.~\ref{fig:q_p_qbar} (left), although the changes are $O(1\%)$ and well within the uncertainty bands. However, the effect on the down singlet (right) is minimal due to its smaller charge, and is largely removed upon refitting. 

The impact on the valence quarks is shown in Fig.~\ref{fig:qv}. The same $q \rightarrow q + \gamma$ emission as before plays the dominant role here; this mimics the impact of QCD DGLAP on the valence quarks, due to gluon emission, i.e. both the quarks and antiquarks are shifted to lower $x$, and hence the valence difference tends to reduce at intermediate to high $x$. This effect is visible in the up valence, which is reduced at intermediate to high $x$, with a corresponding increase observed at lower $x$, due to the valence sum rule (though the size of the impact  at the very lowest values of $x$ is to some extent driven by extrapolation). Again, the impact on the down valence is rather milder.

Finally, the effects of the inclusion of QED effects on the total strangeness, $s+\bar{s}$, and on the gluon are illustrated in Fig.~\ref{fig:s_gluon}. The strangeness will be sensitive to the same photon emission effect as the up and down quarks, and indeed some reduction is observed in the high $x$ region, though not at the highest values of $x$. However, this reduction extends to intermediate and low values of $x$. This is due to the addition of the photon PDF, which carries a fraction of the proton momentum, and hence requires a reduction in the size of the other PDFs in order to satisfy the momentum sum rule. As the strangeness is rather less well constrained than the up and down singlets, it is more affected by this. A similar reduction is observed for the gluon across a broad region, apart from at the very highest values of $x$, where some increase is seen, an effect which seems common to other analyses~\cite{Bertone:2017bme,Xie:2021equ}. 

\begin{figure} [t]
\begin{center}
\includegraphics[scale=0.65]{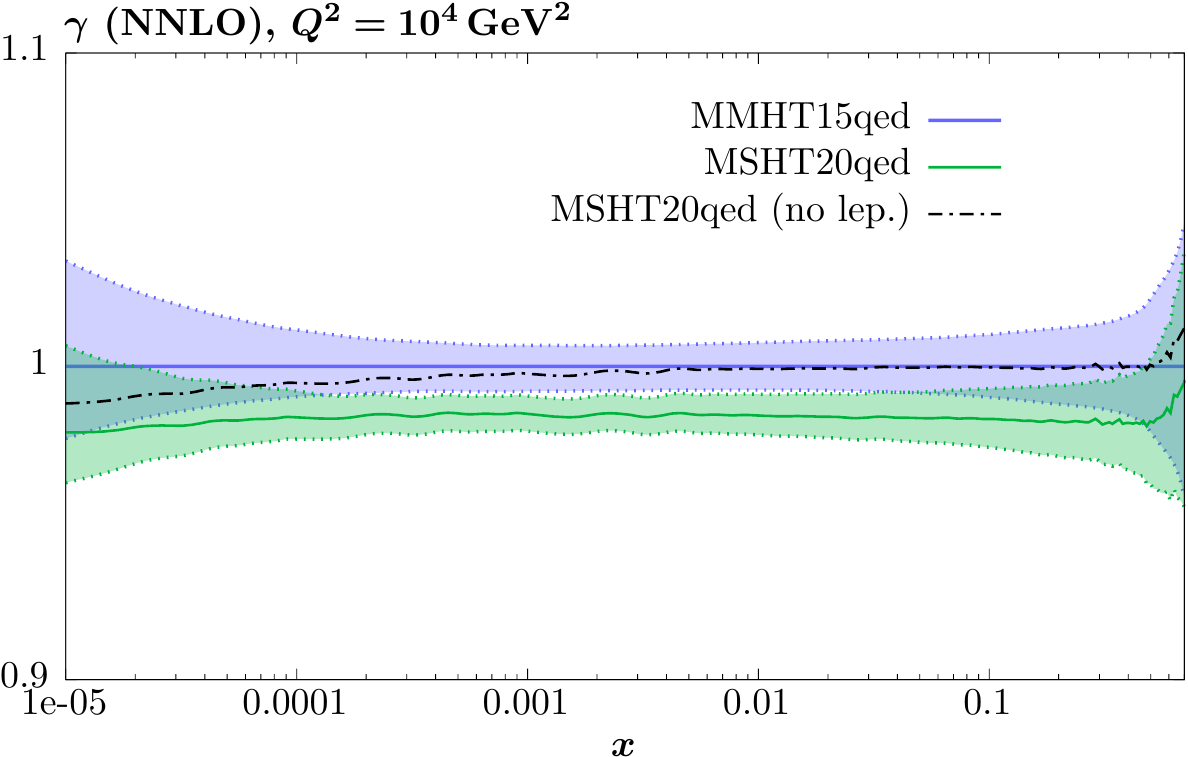}
\includegraphics[scale=0.65]{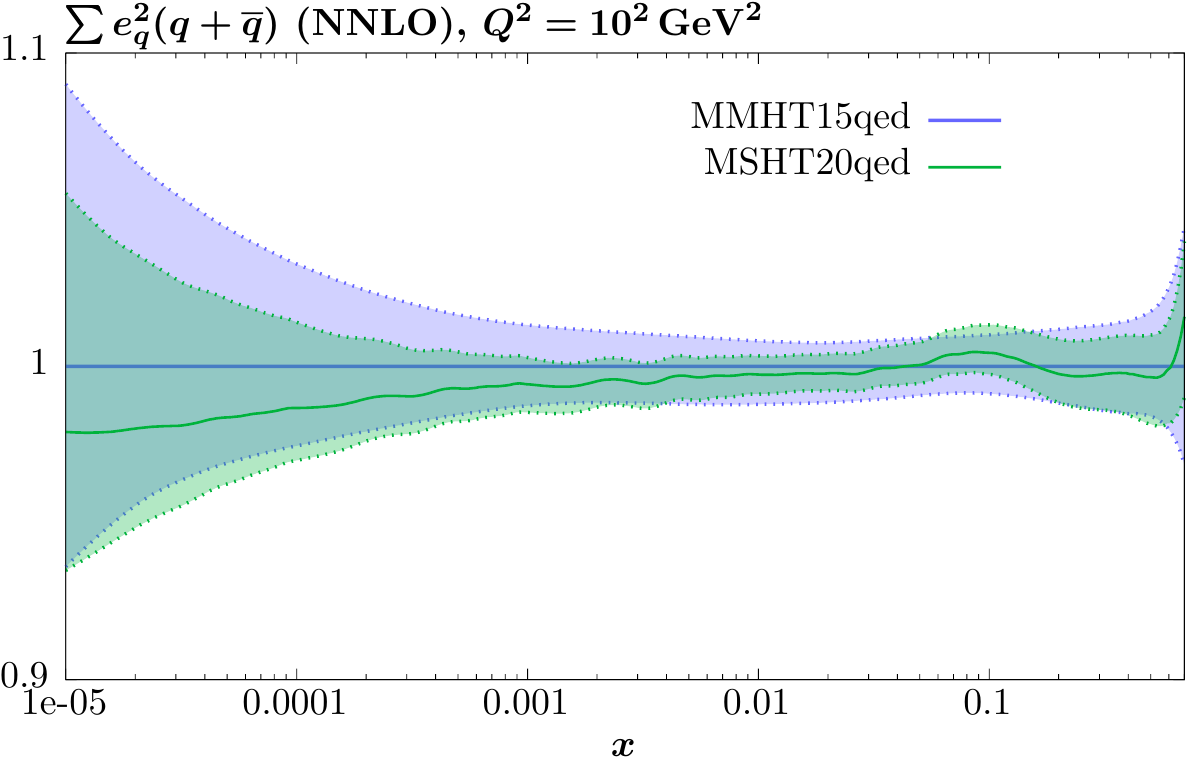}
\caption{\sf The ratio of the (left) photon and (right) charge--weighted singlet distributions (with uncertainties) at $Q^2=10^4$ ${\rm GeV}^2$, resulting from fits to the MSHT20 dataset, to the MMHT15qed case.}
\label{fig:phot}
\end{center}
\end{figure}

\begin{figure} [t]
\begin{center}
\includegraphics[scale=0.65]{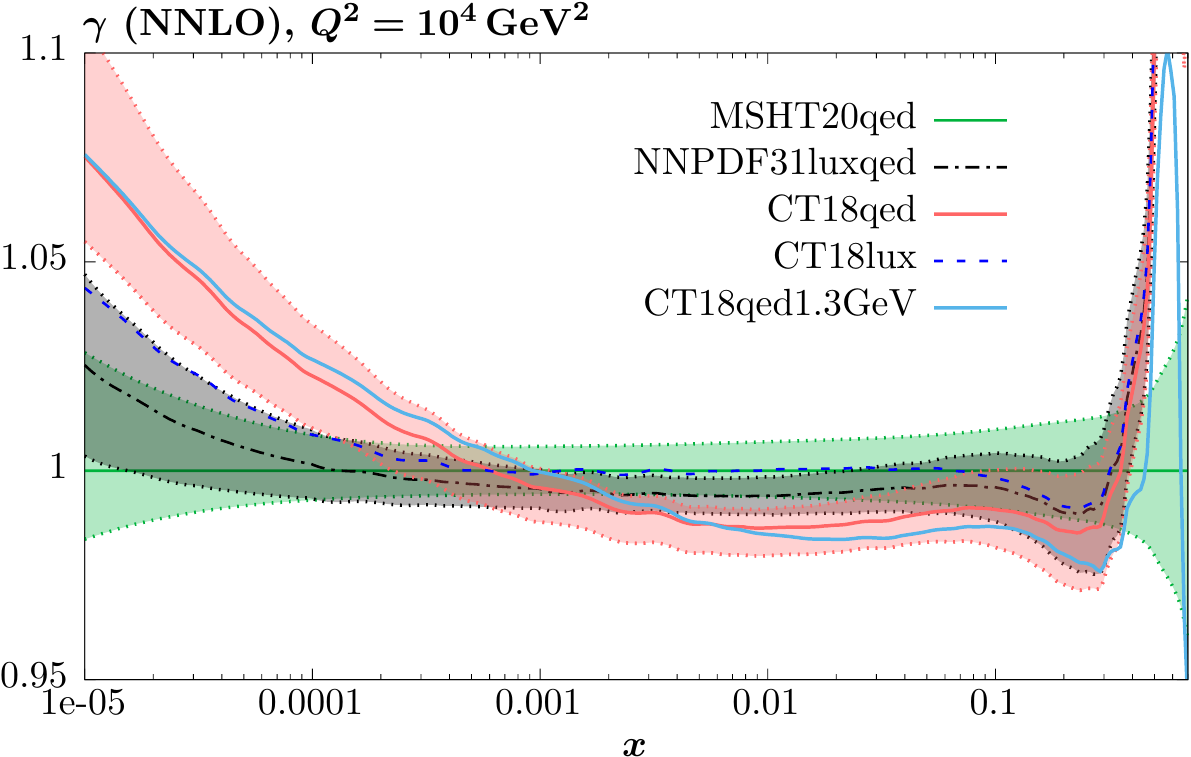}
\includegraphics[scale=0.65]{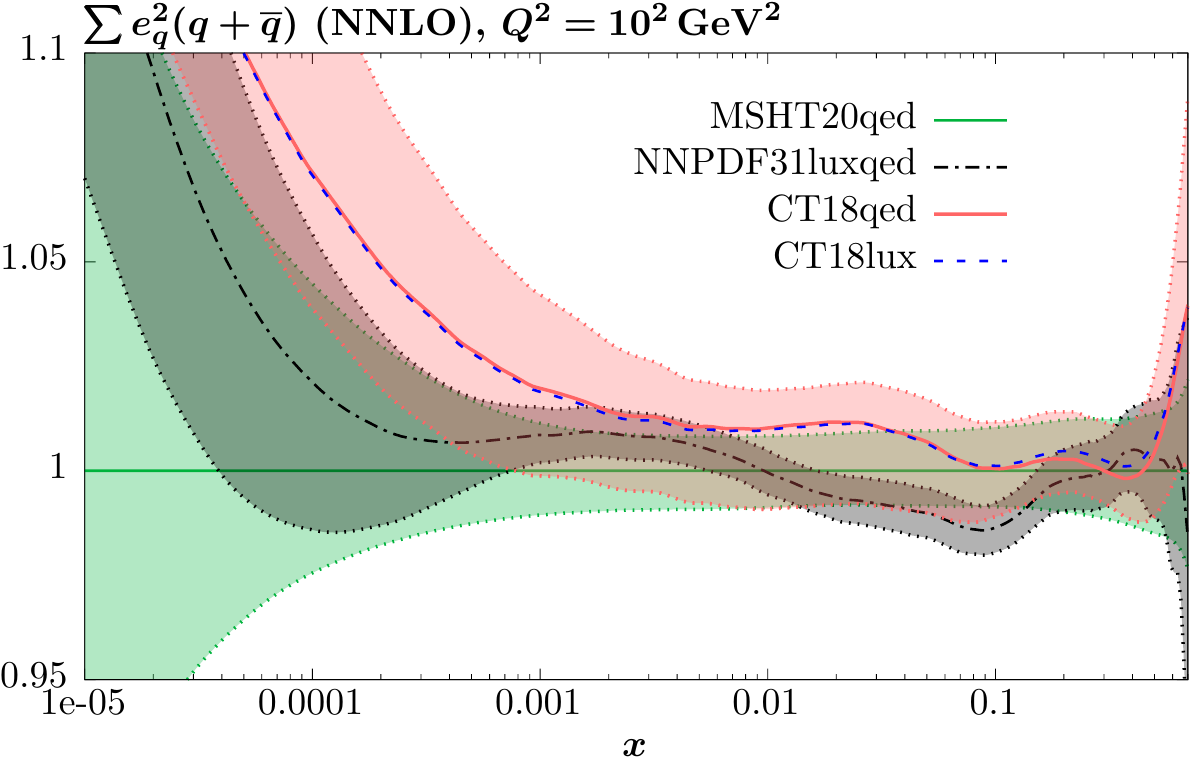}
\caption{\sf The ratio of the (left) photon and (right) charge--weighted singlet distributions for various PDF sets at $Q^2=10^4$ ${\rm GeV}^2$.}
\label{fig:pdfcomp}
\end{center}
\end{figure}

 In Fig.~\ref{fig:phot} (left) we show the change in the photon PDF with respect to the MMHT15qed case. We can see that, as expected from the discussion in Section~\ref{sec:theory}, the photon is now $O(2\%)$ lower across the entire $x$ region. This is almost entirely due to the impact of lepton loops in $P_{\gamma\gamma}$; to demonstrate this we also show the result of performing a fit to the same MSHT20 dataset, but with leptonic loops excluded (i.e. following the procedure of MMHT15qed). We can see that in this case the photon is very similar to the MMHT15qed, apart from at rather lower $x$, where it is somewhat reduced. In Fig.~\ref{fig:phot} (right) we show the charge--weighted quark singlet, which is somewhat lower at intermediate to low $x$, reflecting the difference in the MMHT15 and MSHT20 QCD--only PDFs.  This difference will drive the reduction in the photon at low $x$, due to the reduced contribution from $q\to q + \gamma$ emission. We note that the impact of including lepton loops in $P_{\gamma\gamma}$ on all other partons is very minor, and is for that reason not shown here.
 
In Fig.~\ref{fig:pdfcomp} we compare the MSHT20qed photon PDF with other results in the literature, namely the NNPDF31luxqed~\cite{Bertone:2017bme}, and CT18qed, CT18lux~\cite{Xie:2021equ} sets. These all apply the same basic LUXqed approach as outlined in~\cite{LUXQED1,LUXQED2} and used for the MSHT set, but differ in the specifics of the implementation, as well as the underlying QCD partons. In more detail, the CT18qed set applies a similar modification to us, namely applying the LUXqed formula for the photon at input scale $Q_0$, before evolving with standard QED DGLAP. On the other hand,  NNPDF31luxqed and CT18lux apply the LUXqed formula at higher scales, see~\cite{Bertone:2017bme,Xie:2021equ} for more details. We can see that for intermediate to reasonably high values of $x$ the agreement between the sets is  good, as we might expect. At low $x$ the CT and NNPDF photons lie somewhat above MSHT, which from Fig.~\ref{fig:pdfcomp} (right) we can see is largely driven by the difference in the charge weighted quark singlet PDFs, via their impact on the photon through DGLAP evolution. At the highest values $x\gtrsim 0.5$, on the other hand, the MSHT photon is lower than the other results. In~\cite{Xie:2021equ} it is argued that the MSHT `$Q_0$' approach tends to lead to a lower photon at high $x$ in comparison to the high scale approach, 
due to the difference in treatment of non-leading twist contributions to $F_2(x,Q^2)$ above $Q_0^2$, and hence this could explain the difference with respect to the NNPDF31luxqed and CT18lux sets. We can see that the CT18qed set, which applies the same basic `$Q_0$' methodology as MSHT, remains higher than MSHT (though indeed smaller than CT18lux, even if this is not evident in the plot), but there is better agreement with CT18qed1.3GeV, which has a more similar starting scale to MSHT, whereas the default CT18qed uses $Q_0=3~\GeV$. We note that in~\cite{Xie:2021equ} it was observed that the MMHT15qed set was similar in size to CT18qed. However, as discussed above this excluded leptonic loop contributions to $P_{\gamma\gamma}$ (which are included in CT18), which as observed in Fig.~24 of~\cite{MMHTQED} reduce the photon PDF most prominently at high $x$.
 
 \begin{figure} 
\begin{center}
\includegraphics[scale=0.7]{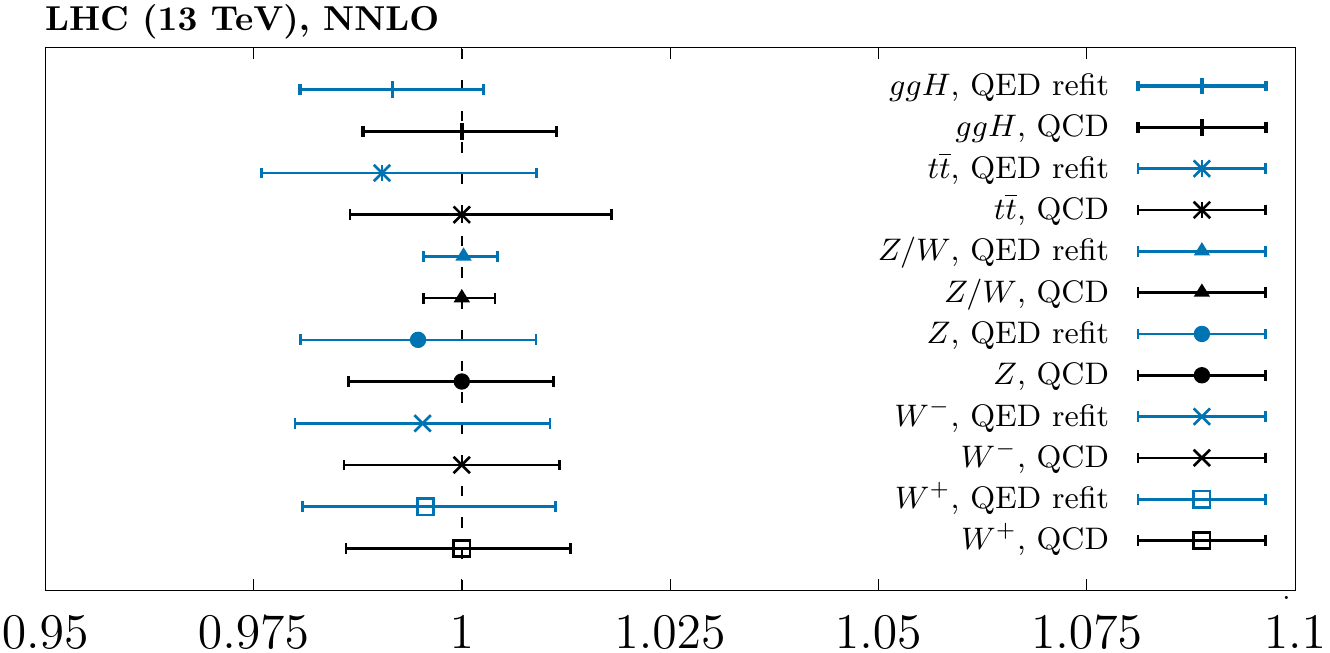}
\caption{\sf Benchmark cross sections obtained with  NNLO fits to the MSHT20 dataset, with QED effects included to that without.. Results are normalized to the central value of the QCD only fit.}
\label{fig:Benchmarks}
\end{center}
\end{figure} 
 
Finally, in Fig.~\ref{fig:Benchmarks} we show results for a range of benchmark cross sections, namely Higgs boson production in gluon fusion, top quark pair, and $W,Z$ production. 
To calculate the cross section at NNLO in QCD perturbation theory we use the same procedure as described in~\cite{Bailey:2020ooq}. That is, we use LO electroweak perturbation theory, with the $qqW$ and $qqZ$ couplings defined by 
\begin{equation}
  g_W^2 =  G_F M_W^2 / \sqrt{2}, \qquad g_Z^2 = G_F M_Z^2 \sqrt{2}, 
\end{equation}
and other electroweak parameters as in~\cite{MSTW}. We take the Higgs mass to be 
$m_H=125~\GeV$  and the top pole mass is $m_t=172.5~\GeV$.  
For the $t\overline{t}$ cross section we use \texttt{top++}~\cite{NNLOtop}. We note that in all of these cases the impact of PI production is very small, and hence is not included; that is, all changes result from the impact on the quark and gluon PDFs as a result of the refit including QED effects.

The results are plotted as ratio of the central value for the pure QCD case. For the Higgs and top quark cases we can see that the QED fit results in central cross section values that are $\sim 1\%$ lower. Although these shifts, which are driven by the lower gluon PDF at intermediate to high $x$ values seen in Fig.~\ref{fig:s_gluon}, are only moderate in size, we note that they are smaller than but of a similar order to the PDF uncertainty on the cross section, and hence are certainly significant in impact with respect to this. The $W,Z$ cross section are similarly reduced in the QED case, albeit by a somewhat smaller amount. This is driven by the reduced quark PDFs in the QED fit. For the $Z$ to $W$ ratio on the other hand, which is relatively insensitive to such normalization effects, the impact is seen to be rather small and well within uncertainties. We can in addition see that the PDF uncertainties in the QED and QCD cases are rather comparable, reflecting the similar uncertainties in the PDFs themselves.

\section{Breakdown between elastic/inelastic components and neutron PDFs} \label{sec:elinel_neutron}

\begin{figure} 
\begin{center}
\includegraphics[scale=0.65]{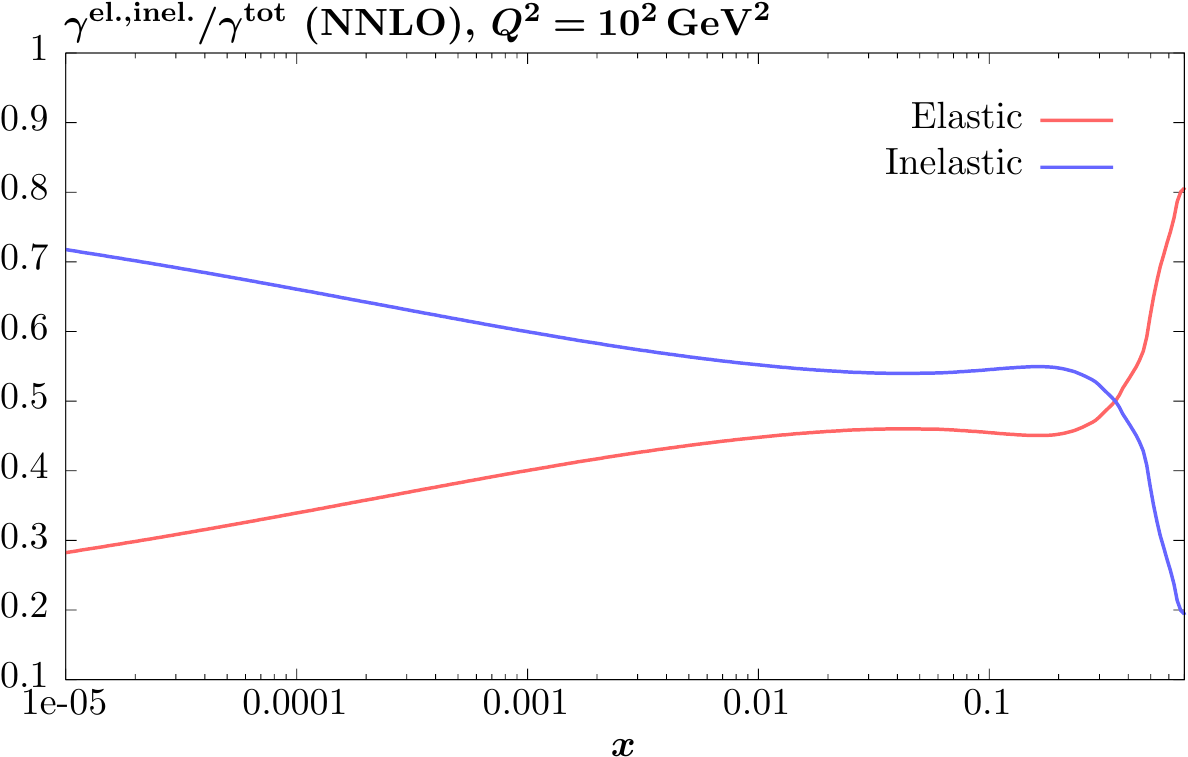}
\includegraphics[scale=0.65]{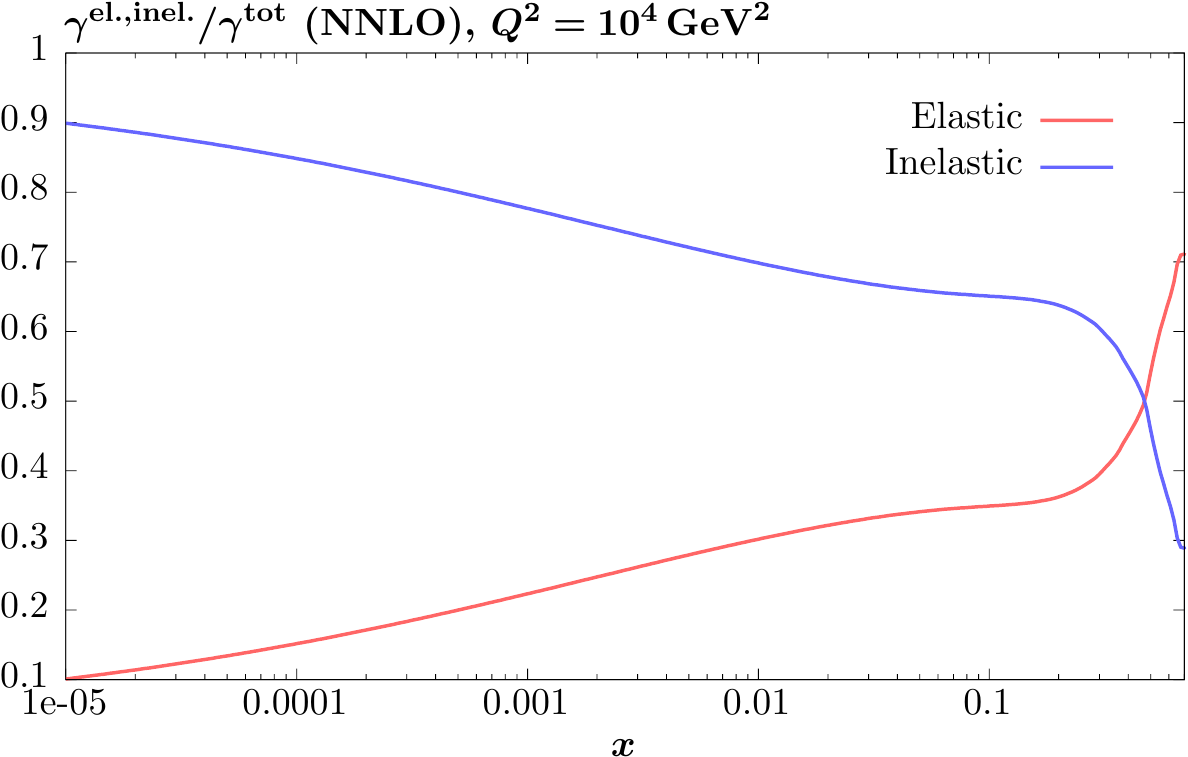}
\caption{\sf  Ratio of the inelastic and elastic components to the total (given by the sum of both) photon. Results at $Q^2=10^2 (10^4)$ ${\rm GeV}^2$ are shown in the left (right) plots.}
\label{fig:phot_inel_el}
\end{center}
\end{figure}

As in~\cite{MMHTQED} we provide the individual elastic, $\gamma^{\rm el.}(x,Q^2)$, and inelastic, $\gamma^{\rm inel.}(x,Q^2)$, photon PDF components for our latest fit, with $\gamma(x,Q^2) = \gamma^{\rm el.}(x,Q^2)+\gamma^{\rm inel.}(x,Q^2)$. This separation is achieved in exactly the same way as described in~\cite{MMHTQED}, to which we refer the reader for details. It can for example be useful when making predictions for exclusive and semi--exclusive PI production~\cite{Harland-Lang:2016apc,Harland-Lang:2020veo}, although in this case care must be taken to also include the survival factor probability of no additional particle production due to multi--particle interactions (MPI), see e.g.~\cite{Khoze:2013dha}. The fractional contribution from these components to the total at different scales is shown in Fig.~\ref{fig:phot_inel_el}. We can see that at $Q^2 = 10^4$ ${\rm GeV^2}$ the inelastic component is dominant until very high $x$. At the lower scale of $Q^2 = 10^2$ ${\rm GeV^2}$ on the other hand the relative contribution from the elastic component is somewhat larger, due to the shorter evolution length for (inelastic) $q\to q \gamma$ splitting. The breakdown is very close indeed to that in the MMHT15qed set, which is not shown for clarity.

\begin{figure} 
\begin{center}
\includegraphics[scale=0.65]{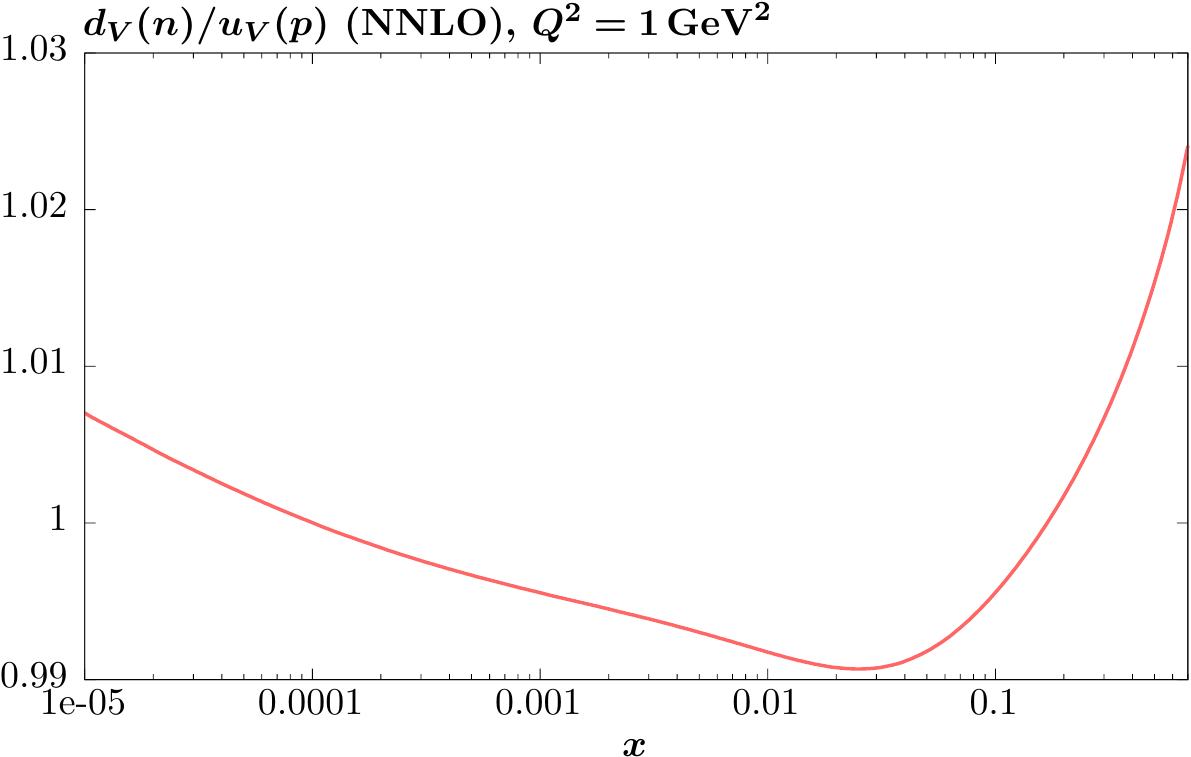}
\includegraphics[scale=0.65]{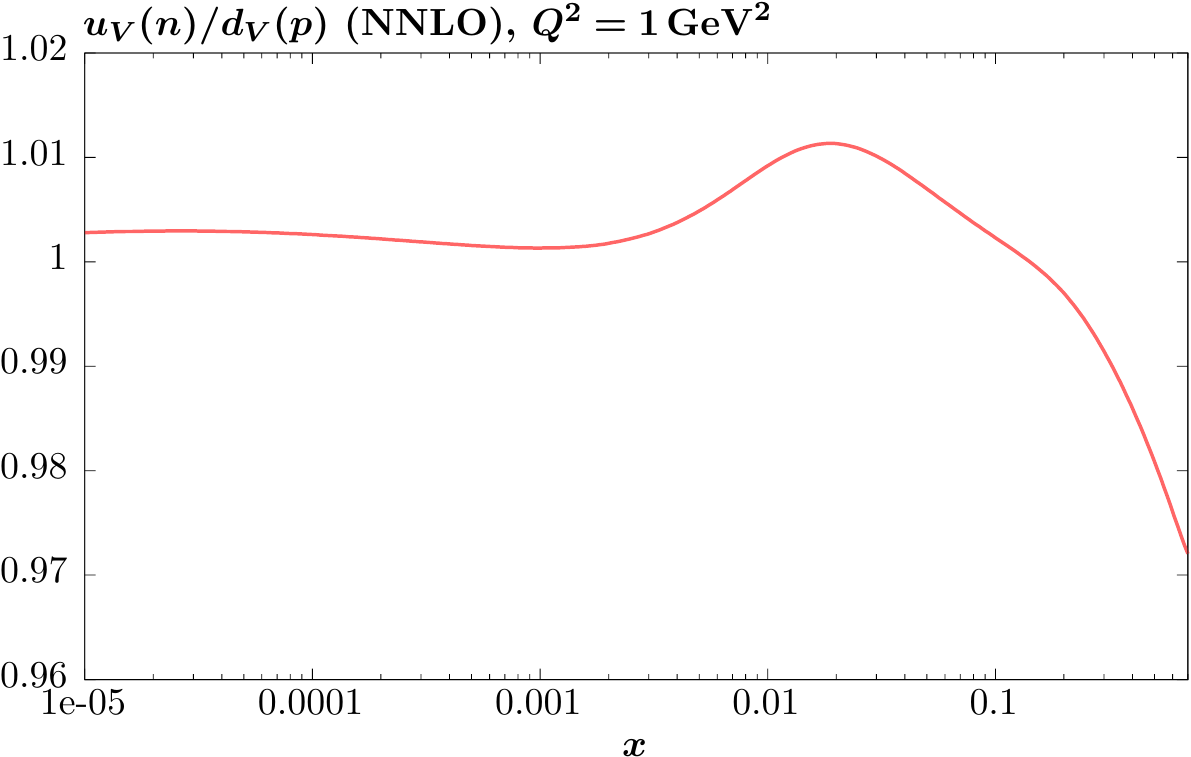}
\caption{\sf  Ratio of neutron down (up) valence to the proton isospin partner, at $Q^2=1$ ${\rm GeV}^2$, shown in the left (right) plots.}
\label{fig:neut_uvdv}
\end{center}
\end{figure}

\begin{figure} [t]
\begin{center}
\includegraphics[scale=0.65]{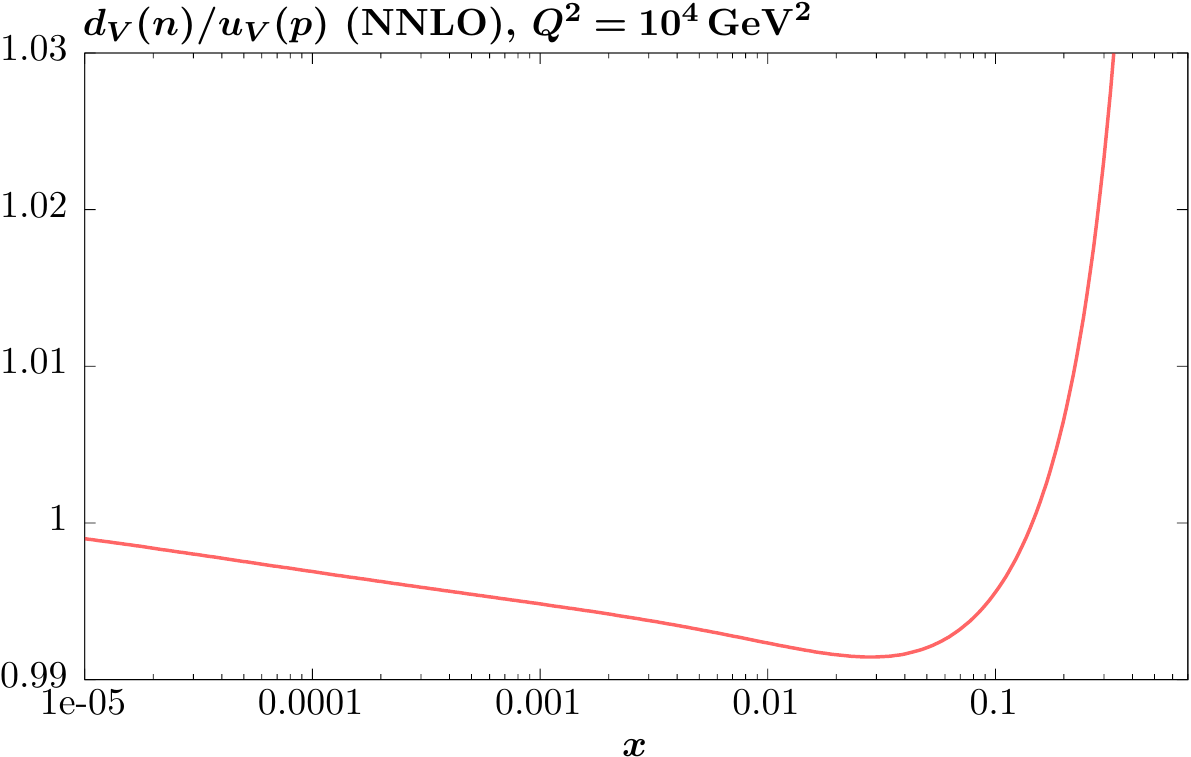}
\includegraphics[scale=0.65]{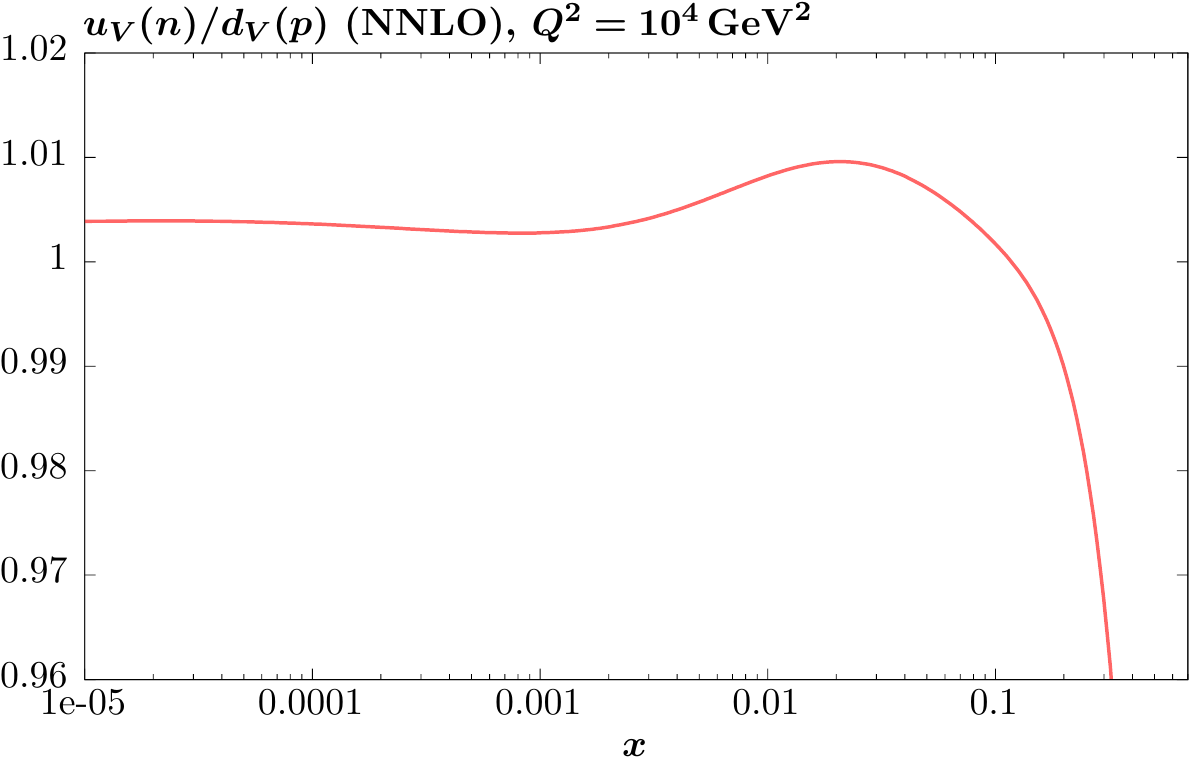}
\caption{\sf  As in Fig.~\ref{fig:neut_uvdv}, but at $Q^2=10^4$ ${\rm GeV}^2$.}
\label{fig:neut_uvdv_q10000}
\end{center}
\end{figure}

We in addition now provide publicly available neutron PDF sets, again following the approach described in the MMHT15qed study~\cite{MMHTQED}. In particular, QED effects are expected to violate the pure assumption of isospin symmetry, for which $d_V(n) = u_V(p)$ and $u_V(n) = d_V(p)$. These will modify the distributions at the input scale $Q_0=1$ GeV, as well as then explicitly in the QED corrected DGLAP evolution to higher scales. The ratio of the neutron down and up valence quarks, at the input scale, to their isospin symmetry partners is shown in Fig.~\ref{fig:neut_uvdv}. We can see that the effect of isospin violation is small, at the 1\% level around the peaks of the valence distributions. These results are comparable to the MMHT15qed case, though interestingly the impact of isospin violation on the $u_V(n)/d_V(p)$ ratio is significantly less at low and high $x$, which is most likely a result of the rather different proton down valence in the MSHT20 fit with respect to the MMHT14 case, due to the more flexible parameterisation as well as the impact of new data in the fit, see~\cite{Bailey:2020ooq} for further discussion. The same comparison, but at $Q^2=10^4$ ${\rm GeV}^2$, is shown in Fig.~\ref{fig:neut_uvdv_q10000}. Broadly, we can see that at high $x$ the neutron $d_V$ ($u_V$) is enhanced (suppressed) with respect to the proton $u_V$ ($d_V$), due to the lower (higher) electric charge of the corresponding neutron PDFs, and hence less (more) significant QED radiation effects, which tend to reduce the valence distribution in this region. As expected from the form of the prescription for the input neutron PDFs, this trend is already present in the distributions at input, and then is clearly enhanced by the effect of QED DGLAP as we evolve to $Q^2=10^4$ ${\rm GeV}^2$.

\begin{figure} [t]
\begin{center}
\includegraphics[scale=0.65]{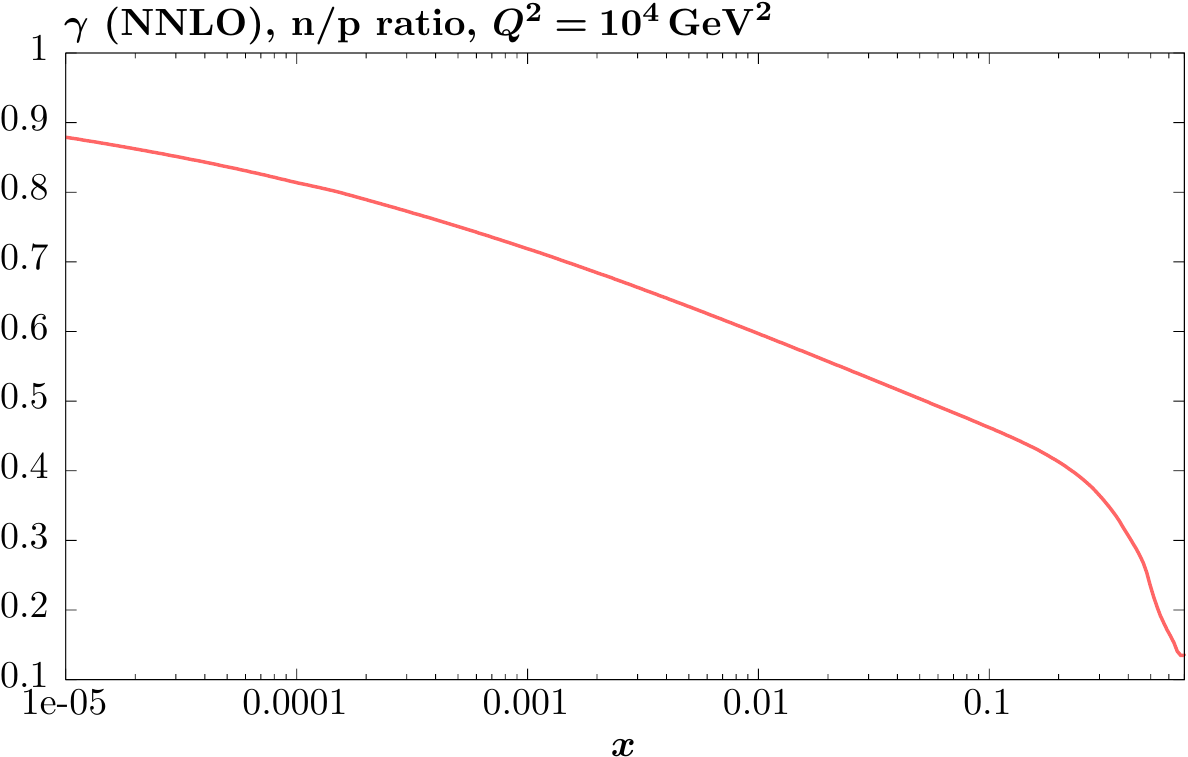}
\includegraphics[scale=0.65]{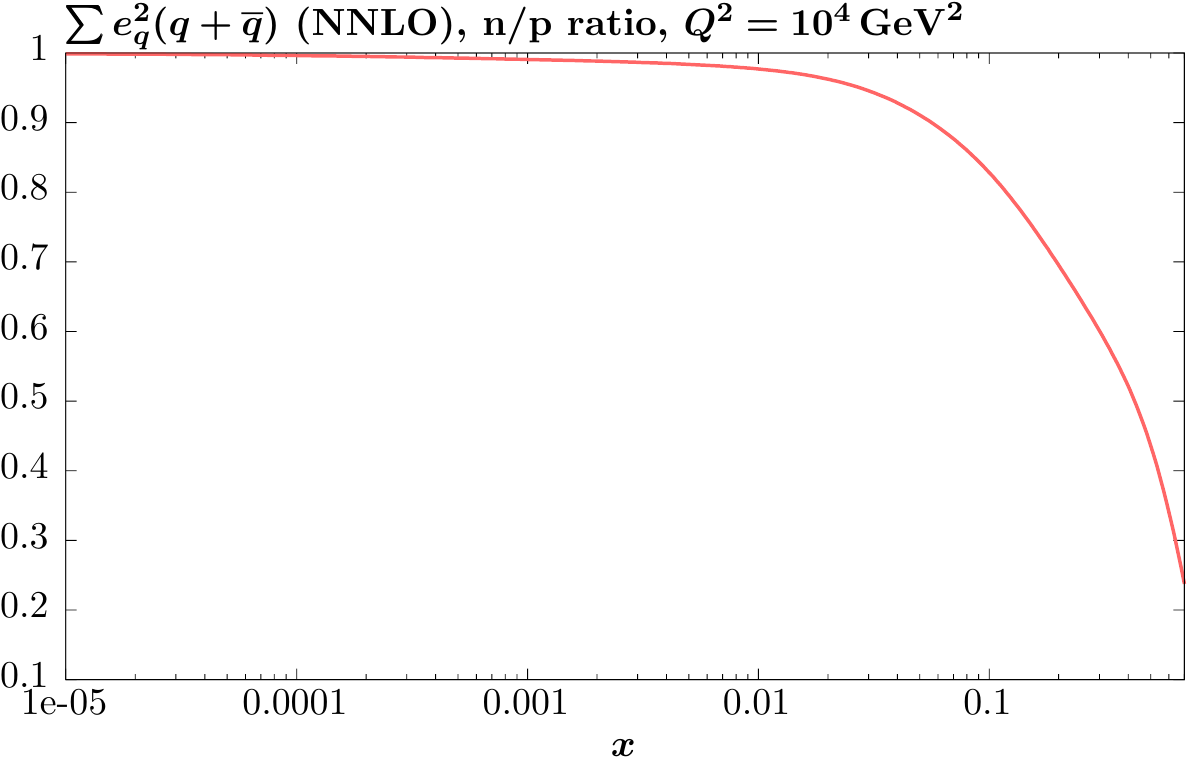}
\caption{\sf  Ratio of neutron photon (charge--weighted singlet) PDFs to the proton case, at $Q^2=10^4$ ${\rm GeV}^2$, shown in the left (right) plots.}
\label{fig:neut_phot}
\end{center}
\end{figure}

Finally, in Fig.~\ref{fig:neut_phot} (left) we show the ratio of the photon PDF in the neutron to the proton case. We can see that the neutron's photon PDF is rather lower than that in the proton, due in part to the significantly smaller elastic component of the photon in this case, but also the suppression in the charge--weighted singlet quark PDF at higher $x$ (Fig.~\ref{fig:neut_phot} (right)), and hence the smaller inelastic photon component that this will generate. At low $x$ the sea quarks dominate and this ratio tends to unity. Hence, at low $x$ the suppression of the  photon PDF in the neutron is observed to be less significant, though this is also due to the smaller relative elastic component in the proton case at low $x$, as seen in Fig.~\ref{fig:phot_inel_el}.

\section{PDF availability}\label{sec:avail}

We provide the MSHT20 PDFs in the \texttt{LHAPDF} format~\cite{Buckley:2014ana}:
\\
\\
\href{http://lhapdf.hepforge.org/}{\texttt{http://lhapdf.hepforge.org/}}
\\
\\
as well as on the repository:
\\
\\ 
\href{http://www.hep.ucl.ac.uk/msht/}{\texttt{http://www.hep.ucl.ac.uk/msht/}}
\\
\\
We present NNLO eigenvector sets of PDFs at the default value of $\alpha_S(M_Z^2)=0.118$:
\\
\\
\href{http://www.hep.ucl.ac.uk/msht/Grids/MSHT20qed_nnlo.tar.gz}{\texttt{MSHT20qed\_nnlo}}
\\
\\
but not at NLO, as it is only at NNLO QCD level of precision that QED corrections become relevant. We also provide  equivalent PDF sets, but with only the elastic or inelastic components of the photon output (see~\cite{MMHTQED} for discussion of how this is achieved):
\\
\\
\href{http://www.hep.ucl.ac.uk/msht/Grids/MSHT20qed_nnlo_elastic.tar.gz}{\texttt{MSHT20qed\_nnlo\_elastic}}\\
\href{http://www.hep.ucl.ac.uk/msht/Grids/MSHT20qed_nnlo_inelastic.tar.gz}{\texttt{MSHT20qed\_nnlo\_inelastic}}
\\
\\
Finally, we also present neutron PDF sets, including the elastic and inelastic photon components:
\\
\\
\href{http://www.hep.ucl.ac.uk/msht/Grids/MSHT20qed_nnlo_neutron.tar.gz}{\texttt{MSHT20qed\_nnlo\_neutron}}\\
\href{http://www.hep.ucl.ac.uk/msht/Grids/MSHT20qed_nnlo_neutron_elastic.tar.gz}{\texttt{MSHT20qed\_nnlo\_neutron\_elastic}}\\
\href{http://www.hep.ucl.ac.uk/msht/Grids/MSHT20qed_nnlo_neutron_inelastic.tar.gz}{\texttt{MSHT20qed\_nnlo\_neutron\_inelastic}}
\\
\\
In all cases these contain 38 eigenvectors, of which 6 are due to uncertainties in the determination of the photon PDF.

\section{Conclusions}\label{sec:conc}

In this paper, we have presented the MSHT20qed NNLO PDF set. This closely follows the MSHT20 global analysis~\cite{Bailey:2020ooq}, but includes QED corrections to the PDF evolution, and a corresponding photon PDF of the proton. The photon PDF is calculated at input following the LUXqed approach, such that percent level PDF uncertainties in the photon PDF are achieved, and a full refit is performed. From this, we have made available a the fully consistent set of QED corrected partons.

We have presented a detailed overview of the  expectations for the relevance of photon--initiated (PI) production in processes that enter current global PDF fits. We have seen that in most cases the PI contribution, calculated using a suitable photon PDF set based on the LUXqed approach, such as MSHT20qed, is found to be very small, entering at the per mille level. These can therefore where necessary be safely included via standard NLO EW K--factors, as part of the broader class of EW corrections, provided a photon PDF set consistent with the LUXqed approach is used. In general however, their impact on the PDF fit is expected to be marginal at the current level of precision, and these are significantly suppressed with respect to other dominant EW corrections, and in particular those due to the presence of Sudakov EW enhancements. 

On the other hand, the contribution from the PI subprocess to lepton pair production below and above the $Z$ peak at the LHC can be at the level of $\sim 10\%$ of the standard DY contribution. This is therefore rather distinct from the PI corrections to other processes in the fit, and certainly an accurate and precise account of these is mandatory.  We have achieved this by making of the structure function (SF) calculation, which provides percent level precision in the underlying cross section for this process.

As in the previous MMHT15qed study~\cite{MMHTQED}, we observe some deterioration in the fit quality with respect to a QCD--only fit. However, this effect is relatively mild, and certainly the inclusion of not only PI channels in the fit, where relevant, but also the QED corrections to the PDF evolution of the QCD partons, and their subsequent impact on the PDF sets through refitting, will provide a more accurate result. We have compared our results to a pure QCD--only fit and have found that the impact on the quark and gluon PDFs can be non--negligible, though it is always currently within the PDF uncertainties of the QCD set. It should be noted though that there is in principle no requirement that this should be the case, given that the PDF uncertainties are of a distinct origin to such differences. This is also seen when considering a range of benchmark cross sections: predictions for Higgs boson production via gluon fusion, $W,Z$ production and $t\overline{t}$ production at the 13 TeV LHC change at the level of 1\%, which is smaller than, but in some cases comparable to the underlying PDF uncertainty.

We provide a NNLO error set for public use, as well as the breakdown between the photon elastic and inelastic components, and the corresponding neutron PDF set, including QED driven isospin violation. Such QED--corrected PDF sets will play a key role in future LHC precision phenomenology.

\vspace{0.5cm}

\section*{Acknowledgements}

We would like to thank numerous members of the PDF4LHC committee and working 
group for useful conversations. 
T. C. and R. S. T. thank the Science and Technology Facilities Council (STFC) for support via grant awards ST/P000274/1 and ST/T000856/1. L. H. L. thanks STFC for support via grant award ST/L000377/1.

\newpage

\bibliography{references.bib}

\bibliographystyle{h-physrev}

\end{document}